\documentclass[english,12 pt]{article}
\usepackage[T1]{fontenc} 
\usepackage[latin9]{inputenc}
\usepackage{graphicx}
\usepackage{babel}
\usepackage{amsmath}
\usepackage{multicol}
\makeatletter

\newcommand{\fig}[1]{Figure (\ref{#1})}
\newcommand{\eq}[1]{Eq. (\ref{#1})}
\newcommand{\eqs}[1]{Eqs. (\ref{#1})}
\newcommand{\se}[1]{Section (\ref{#1})}

\newcommand{\la}[1]{ \label{#1}}
\renewcommand{\a}{\alpha}
\renewcommand{\b}{\beta}
\newcommand{\e}{\epsilon}
\renewcommand{\r}{\mathbf{r}}
\newcommand{\x}{\mathbf{x}}

\newcommand{\R}{\mathbf{R}}
\newcommand{\s}{\mathbf{s}}

\renewcommand{\d}{\partial}

\newcommand{\bsubs}{\begin{subequations}}
\newcommand{\esubs}{\end{subequations}}

\newcommand{\be}{\begin{equation}}
\newcommand{\ee}{\end{equation}}

\newcommand{\bea}{\begin{eqnarray}}
\newcommand{\eea}{\end{eqnarray}}

\begin{document}

\title{Relating Theories via Renormalization  
% \\DraftV4,0
}
\author{ Leo P. Kadanoff\\
The James Franck Institute\\
The University of Chicago\\
Chicago, Illinois, USA\\
and\\
The Perimeter Institute\\
Waterloo, Ontario, Canada\\
email: leop@UChicago.edu
\footnote{Copies of this paper will be published on arXiv and on the author's web site:  jfi.uchicago.edu$\backslash$$\sim$ leop$\backslash$. }}
\maketitle
\begin{abstract}
The renormalization method is specifically aimed at connecting theories describing physical processes at different length scales and thereby connecting different theories in the physical sciences. 

The renormalization method used today is the outgrowth of one hundred and fifty years of scientific study of thermal physics and phase transitions.  Different phases of matter show qualitatively different behavior separated by abrupt phase transitions. These qualitative differences seem to be present in experimentally observed condensed-matter systems.   However,  the ``extended singularity theorem'' in statistical mechanics shows that sharp changes can only occur in infinitely large systems. Abrupt changes from one phase to another are signaled by fluctuations that show correlation over infinitely long distances, and are measured by  correlation functions that show algebraic decay as well as various kinds of singularities and infinities in thermodynamic derivatives and in measured system parameters.   

Renormalization methods were first developed in field theory to get around difficulties caused by apparent divergences at both small  and large scales.  However, no renormalization gives a fully satisfactory formulation of field theory. 

The renormalization (semi-)group theory of phase transitions  was put together by Kenneth G. Wilson in 1971 based upon ideas of scaling and universality developed earlier in the context of phase transitions and of  couplings dependent upon spatial scale coming from field theory.  Correlations among regions with fluctuations in their order underlie renormalization ideas.  Wilson's theory is the first approach to phase transitions to agree with the extended singularity theorem.  

Some of the history of the study of these correlations and singularities is recounted, along with the history of renormalization and related concepts of scaling and universality.  Applications, particularly to condensed-matter physics and particle physics, are summarized.       

This note is partially a summary of a talk given at the workshop ``Part and Whole'' in Leiden during the period March 22-26, 2010. 
\end{abstract}
\tableofcontents
%\newpage{ }

\section{Fundamental and Derived Theories}
\subsection{Accomplishments of physics}

The subject of physics has been remarkably successful. The last hundred and fifty years has brought many accomplishments.  Here, I shall focus upon two of these.  First, our view of the various forces in nature has been, to a large extent, unified. This process started when Michael Faraday and James Clerk Maxwell combined the concepts of  electricity and magnetism to form  a single unified theory.   It has continued to this day, culminating in the construction of  the standard model covering the strong, weak, and electromagnetic interactions. 

A second, and perhaps even more remarkable, accomplishment is that physics, working together with chemistry and various materials science disciplines, has explained very many of the qualitative properties of the matter around us.   The theoretical part of this understanding has been achieved by the use of an approach based upon quantum mechanics and electromagnetic forces, but also, as appropriate, by building models based upon classical mechanics and/or phenomenological theories.    This range of theoretical approaches has described a tremendous diversity of physical situations.   Different materials and different thermodynamic phases of the same material have been understood by a collection of models, mostly based upon describing various  kinds of ordering. 

   Taken together, these two kinds of accomplishments provide a vision of the wholeness of the natural world, both in its complexity and its unity.\footnote{Despite this overall complementary structure,  advocates of these two goals have often been in contention.  The argument is partially about the allocation of resources and social {\em cachet}. The  discussion turns on which discipline is more ``fundamental'' or more ``important'' or more ``valuable''.  These kinds of arguments are familiar, but somewhat embarrassing,  parts of the history of science\cite[pp. 45-60]{BrushTemperature}.   } 

\subsection{Relation among different length scales}
Questions about theoretical unity center on how theories that are specific to different length scales end up being consistent with one another.  In condensed matter,  the basic theory  is phrased at the atomic level in which typical lengths are of the order of a tenth of a nanometer.  The goal, then, is to tease out of the atomic theory an understanding of what we see at our human scale of one meter or thereabouts. Many material scientists work at understanding connections between the macro and micro worlds.   This process is, in part, accomplished by setting up macroscopic theories like fluid mechanics, elasticity, thermodynamics, and acoustics as descriptors of the macro world and then seeing how these might be obtained from more microscopic theories.  Science has organized itself by having different people and different disciplines study the various kinds of materials: gases, liquids and solids;  electrical conductors, semiconductors, and insulators; ceramics, metals, and composite materials; ferromagnets, liquid crystals, ..., on and on.    

In this year, 2010, one pressing issue in particle physics is how the standard model, which has as its typical distances $10^{-15}$ meters, can be understood in terms of theoretical constructs like quantum gravitational interactions,  which have a characteristic scale of $10^{-35}$ meters, or in terms of the putative scale for the unification of the strong and electroweak interactions, $10^{-33}$ meters, or perhaps at the terascale (roughly $10^{-18}$ meters) under study at the Large Hadron Collider.   Different length scales require different theories\footnote{Some physicists even hope to find a ``final'' theory with the potential for working at all possible scales.  They then hope that this final theory will serve as the basis for more specialized theories that each work in a limited domain.}.   Our problem is to understand the connections among these different theories.   A particularly vexing and difficult challenge is to fit together gravity, working on the largest scales, with quantum theory, working on the smallest.

\subsection{Connections among theories: Reduction}
The philosophy of science has long struggled with the issue of how different theories, perhaps at different length scales,  might be interconnected. In a  classic view of Ernest Nagel\cite{Nagel}, described by the phrase ``theory-reduction,''  a more  fundamental theory will imply a less fundamental one as a limiting case. In this situation, the  laws and relations of the latter theory are then  a logical consequences of the laws and relations of the more fundamental one.      A familiar example is the one in which Galilean relativity\cite{Galileo} is a consequence of Einstein's special relativity\cite{railway} specialized to situations in which all speeds are small compared with the speed of light. 

The logic of this example is clear.    One can indeed derive Galilean relativity from Einstein's theory. Of course this example, and most of the other examples of theory reduction,  have the difficulty that the described arrow of implication reverses the actual direction of historical development.   For example, Einstein explicitly used Galileo's arguments in his discussion of the logical basis of special relativity. 

The particular  example that might be germane to this paper is the combined history of ``thermal physics'', which can be seen as a combination of kinetic theory, thermodynamics, and statistical mechanics.  This example has been extensively analyzed by Stephen G. Brush, who concludes\cite[pages 524-526]{BrushGases} that the historical connections among these disciplines is so complex and multi-directional as to make the concept of theory reduction worthless for this case.  I concur with his conclusion.   

\subsection{Connections among theories: Analytic continuation\la{continuation}}
The mathematics of analytic continuation provides another fundamental method of connecting theories. Analytic continuation can push known mathematical functions into new territory. Given a function, $f(x)$, that is sufficiently smooth and has known behavior in some region of its argument, it is then possible to find out the behavior of $f$ in other, nearby, regions of possible $x$-values.  The idea is often applied to functions of a complex variable.  The simplest and most important application of analytic continuation is the connection between theories in four dimensional Euclidean space, where the distance between points $x$ and $y$ is $$
d=\big[\sum_{\a=1}^4 (x_\a-y_\a)^2\big]^{1/2}
$$  
and theories in Minkowski space where the appropriate measure of spacing is given by
$$
d^2=\sum_{j=1}^3 (x_j-y_j)^2 - (x_0-y_0)^2
$$
Notice that in the Euclidian domain distances are always positive real numbers, while in the Minkowski domain a time-like ``distance'' will be pure imaginary.  

Many physical behaviors can be studied in one of these two spaces and then have the mathematical consequences of the study transferred to the other space via analytic continuation.  The transformation that takes us from one of these spaces to the other is called a {\em Wick rotation}, and is in effect a rotation of the time axis in the complex plane from a real to a pure imaginary direction. 

A  philosopher will immediately notice that concepts do not always survive analytic continuation.  For example, in Minkowski space we can describe different kinds of event-separations  in terms of the light cone; but that has no meaning in Euclidian space.   

For the purposes of this paper, the most important analytic continuation is, in fact, the Wick rotation and its application to statistical mechanics and quantum field theory.  Simply stated, a Wick rotation converts a $d$- dimensional problem in statistical mechanics to a problem in quantum field theory in $d-1$ dimensions of space and one dimension of time.   I do not intend to describe this connection in detail here.  A full exposition can be found in a two-volume study by Claude Itzykson and Jean-Michel Droffe\cite{ID}.   The major physical point is that the techniques of statistical mechanics and of quantum field theory can be very closely related to one another.  Thus people who work in the two areas can learn a lot from one another. It does not follow, however, that they have a full similarity of outlook. 

The strategy followed in this paper is to describe the statistical mechanics in a rough way including some of the mathematics used in the formulation, and then to describe the result of  the applications to quantum field theory without describing the field-theory method in much detail.   I follow that strategy in part because statistical mechanics is my own ``home ground'' and I can describe it with much more authority than I can describe particle physics.

One important property of statistical mechanics is the existence of phase transitions, abrupt qualitative changes in behavior.  These changes occur as a result of modifications of parameters describing the material, like temperature and pressure.  Phase transitions also exist in quantum field theories, and are of considerable interest to particle theorists and cosmologists.   Because these transitions can be most closely studied and observed in the context of condensed-matter physics, we might expect that  understanding will be initially gained mostly  from condensed-matter systems. In the usual course of events, the understanding gained in this way will then be applied to more ``fundamental'' branches of physics.  However, in contexts in which basic knowledge is obtained first from the study of earthbound materials and then applied in more ``far-out'' domains,  it is hard to say which kind of scientific work is fundamental, and which applied. 

In recent years, the deepest understanding of phase transitions has been obtained via renormalization techniques.   These methods were developed first in classical field theory\cite{Lorentz,DiracBS} (i.e. classical  statistical mechanics), extended to quantum field theory\cite{Kramers,DysonPT,Schweber}, brought to maturity in application to phase transitions\cite{Wilson}, and then triumphantly reapplied to quantum field theory\cite{HVa,HVb, Politzer,GW}.     

\subsection{Theoretical riches\la{riches}}
Particle physics {\em per se} is not a rich subject.   There are four kinds of forces and only a few different basic theories.  To see riches, one must look to applications of particle physics in astrophysics and cosmology, where one can find a huge variety of astronomical objects.  In a similar vein, the theoretical sources of condensed-matter physics are very limited indeed:  electromagnetic forces, quantum theory, a little relativity.  However, the manifestations of these sparse ingredients are amazingly rich.  

There is a wonderful variety in the different kinds of ordering apparent in materials.
Every physical property can be the basis of several different kinds of molecular or atomic ordering, each kind of order extending over the entire, potentially infinite, piece of material.  The properties that can be ordered include variations in the local density in a liquid-gas transition, frozen waves of density variation in a liquid crystal,  the lining up of spins in a ferromagnet, the spatial extension of electron wave functions and orbits in metals and semiconductors, localization of orbits in insulators and semiconductors, the size of connected clusters in a percolation transition, the alignment of molecules in a ferroelectric,   condensed quantum wave functions in a superfluid or superconductor, $\dots$.

This diversity has been seen in laboratory experiment and observations of nature. In fact, a walk through almost any natural landscape can convince a thoughtful person of the diversity of natural order.   We see this diversity even within a single material.  For example,  we can observe ice in contact with liquid water or see water vapor (steam) come up from a pot of heated water. Different forms of matter have qualitatively different properties.  Walking on ice is well within human capacity, but walking on liquid water is proverbially forbidden to ordinary humans. There are, in fact, at least eight distinguishable forms of ice.  The distinctions among the different phases of matter have been apparent to humankind for millennia, but only brought within the domain of  scientific understanding since the 1880s.  

The diversity of thermodynamic phases is most clearly manifested in the course of phase transitions. As the thermodynamic phase of the materials change,  different phases of matter are in contact with one another.  Quantitative theories of phase transitions were first introduced via the work of Johannes van der Waals and Maxwell.  Our knowledge deepened through the accomplishments of Lev Landau, Philip Anderson,  Michael Fisher, and Kenneth G. Wilson.    As a result of the efforts of these people and many, many others, it can now be said that  we  have a qualitative understanding of properties of the many materials in our environment.  

In past papers\cite{I,LPKwebsiteII},  I have described the history of phase transition theory by focusing on macroscopic properties of the materials involved,  and indeed mostly upon infinities in their thermodynamic derivatives. These papers are denoted respectively as I and II.  Here, I intend to discuss renormalization theory and phase transitions by looking at spatial variations in the physical properties of the materials.   For example, near a liquid-gas phase transition, the density will vary in such a way that one rather large region of the material can have properties appropriate for a gas while another large region looks more like a liquid. Such variations are called fluctuations. They are particularly apparent near what are called {\em continuous phase transitions.} These large-scale fluctuations produce correlations  over long distances within the material. Physicists use descriptors called {\em correlation functions} to describe localized fluctuations and their correlations over large distances.

The correlation functions will play an important role in this paper. Like infinities in thermodynamic derivatives, these functions  describe the singularities that underlie  phase transitions. Singularities are also important in field theory and particle physics, where they produce the infinities that first drove physicists to think about renormalizations as the basis for describing the properties of observed particles. 

Because these singularities have effects that are spread out over large regions of space we describe them by speaking of ``extended singularities''.  (See paper II, \cite{LPKwebsiteII}, for a discussion of the ``extended singularity theorem'', which describes how such singularities are essential to the understanding of phase transitions.)  In this paper, we shall focus upon the description of these singularities in terms of fluctuations in local thermodynamic properties.

\subsection{Distinctions among theories: Ordering\la{ordering}}
Back to condensed-matter systems. Different kinds of ordering require different physical theories.  In one example,  a crystalline solid has a regular placement of atoms. The possible types of ordered patterns are defined and delimited by the mathematical discipline called {\em group theory}.  Crystals have their own particular behavior manifested in their special kinds of defects termed, e.g.,  color centers,  dislocations, twinning,  glide plains, etc.  Special disciplines, including crystallography and mineralogy, have grown up to describe crystals.  Crystalline behavior has thus engendered its own theories, different from the theories describing more disordered materials like gases or glasses.

More generally, this paper will consider each form of ordering to define its own  theory. In modern condensed-matter theory, ordering can be described via a renormalization calculation, with each kind of order being described as a particular {\em fixed point} (see \se{fixed} below). Each fixed point to define its own kind of theory, connected with its own kind of special ordering.   

There are two types of ordering defined by the two qualitatively different kinds of phase transitions---   first-order transitions and continuous transitions. The distinction is traditionally made in terms of the thermodynamic function called the free energy (see \se{thermo} below).  One variety of free energy depends upon the thermodynamic variables:  temperature, volume, and the number of particles of various kinds.  If the first derivative of the free energy is a discontinuous function of these variables, the phase transition is then said to be of {\em first order}.  Any other kind of abrupt change is described as a {\em continuous phase transition.}   The first-order transition is a jump between different phases, and each phase usually includes a different ordering. The transition in which a solid turns into a liquid by melting is a typical example of a first-order phase transition.  The solid has a crystalline lattice that picks out special directions for its crystal axes; the liquid has no such special directions.   In contrast, the liquid-gas  transition, in which a liquid turns into a gas by boiling, is somewhat less typical in that the two phases involved have the same symmetry properties.   

The continuous phase transition is often seen as a limiting case of a first-order transition in which the jump between the different phases goes to zero. In a fluid, for example water, the difference in density between liquid and gas tends to vary as the temperature is changed. When that difference goes to zero, we say that we have reached the {\em critical temperature.}   In water, when the temperature is increased beyond its critical value, there is no phase transition and no distinction between liquid and gas.

\subsubsection{Kinds of ordering: The liquid crystal example}
Present-day materials science deals with a considerable number of different kinds of orderings and of phases.  We illustrate this by showing a few examples from one small corner of this field, the part that deals with the materials called {\em liquid crystals}.  These are fluids, but unlike everyday fluids like tap water, liquid crystals have long-range order in the arrangement or orientation of their molecules\footnote{ In contrast, in ordinary fluid water, any correlations between the placement and  orientation of the molecules falls off quite rapidly with distance.}.    This order of liquid-crystalline material makes for interesting behaviors in flow and in light scattering.   Controllable light reflection is used  in display devices, for example on the faces of ordinary wristwatches.

Liquid crystals have a variety of different orderings.  The simplest kind is a nematic liquid crystal (see the first panel in \fig{liquidCrystals}).  This material is composed of elongated molecules that tend to line up with one another.  The alignment is seen to be maintained over distances of thousands of times longer than the typical distance between molecules.   This coherent alignment can then be used as a tool in light scattering from the material. 

\begin{figure}
%\begin{multicols}{2}
\includegraphics[height=4cm ]{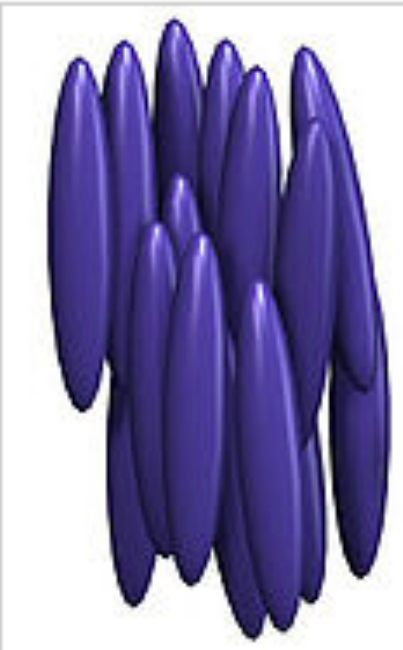}
\hskip 3pt
\includegraphics[height=4cm ]{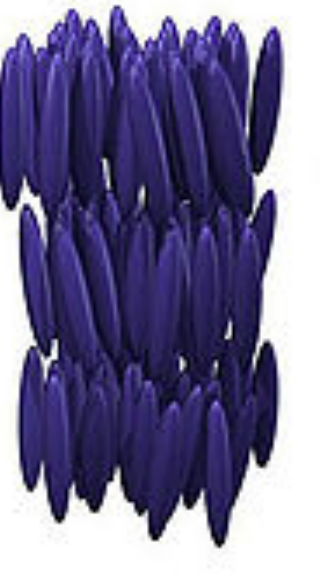}
\hskip 3pt
\includegraphics[height=4cm ]{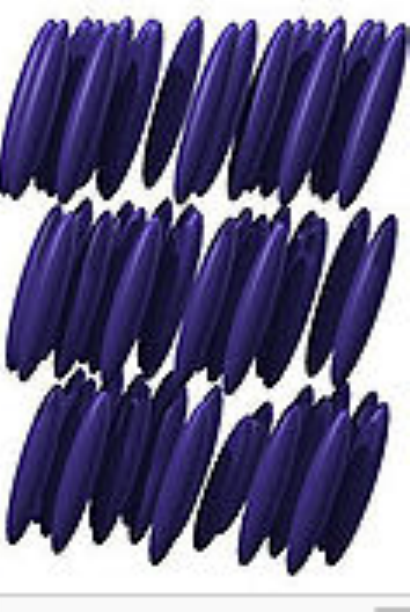}
\hskip 3pt
\includegraphics[height=4cm ]{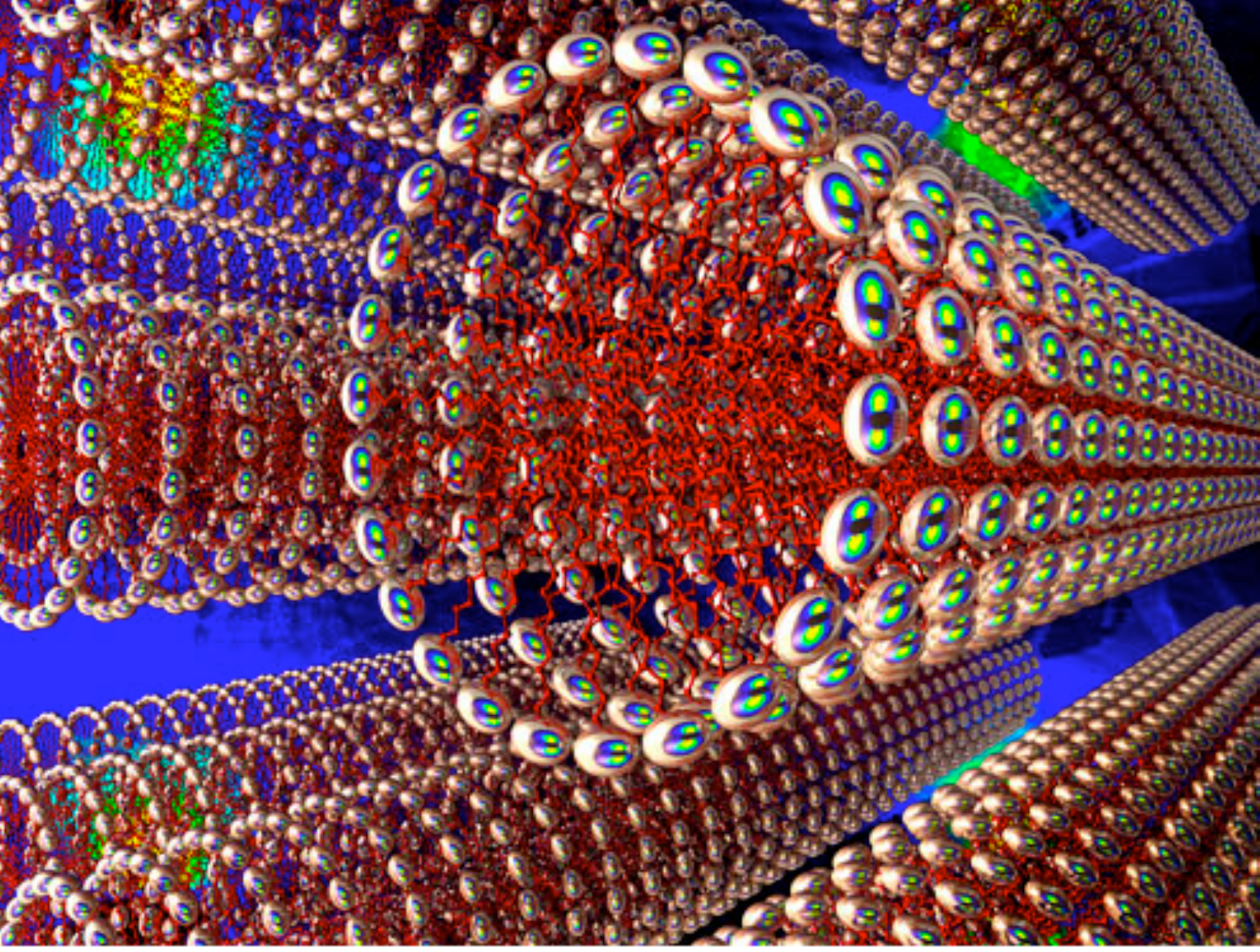}

%\vskip 30 pt
\caption{Four forms of ordering in liquid crystals are illustrated.  The kinds of materials shown in the first three panels are called respectively, from the left,  {\em nematic},  {\em smeticA}, and {\em smeticC} liquid crystals.  In the right-hand panel we see a circular ordering into a form called a micelle.  Many other forms of liquid-crystal-ordering have been observed. }
\la{liquidCrystals}
%\end{multicols}%{2}
\end{figure}

The second panel in the figure shows a material in which the molecules line up, but they also form layers that stack on on top of the other.  The layers then retain their orientation over very long distances. The third panel shows yet another kind of liquid crystal in which once again the molecules line up and form layers but now the alignment is  not perpendicular to the layers, but rather they are tilted.  The right-hand panel shows a structure called a micelle in which the molecules form tubes that run parallel to one another. This kind of structure is found, for example, in many shampoos.

\subsection{Phase transitions and fluctuations}
A fluid with a fixed number of particles is described at the most basic level by giving  the relation among its pressure, $p$, volume, $V$, and temperature, $T$. Andrews\cite{Andrews} used these variables in describing the qualitative properties of carbon dioxide, in particular its liquid-gas phase transition. 

For a range of higher temperatures, the fluid showed a smooth behavior in which the pressure decreased smoothly as the volume of the fluid's container was gradually increased.  For these temperatures,  no liquid-gas phase transition occurred.   
However, at a particular temperature, this behavior changed abruptly.  Below this {\em critical temperature}, the fluid exhibited the familiar phenomenon of boiling.   For a given fluid, and for each temperature, boiling is something that happens at just one value of the pressure. For all other pressures, the volume increases smoothly as the pressure is decreased.      At this special pressure the fluid can equally well support two different forms or {\em phases}.  The lower density phase is called {\em vapor} or {\em gas}; the higher density one is called {\em liquid}.  The boiling is an example of a  first-order phase transition.   

A cook does not see boiling as a discontinuous jump in density.  As the fluid  boils it stops being homogeneous,  but instead is composed of regions differing in their densities, with some regions at the density of vapor and other regions at the density of liquid.  These large regions of the two different densities are jumbled up with one another in an irregular and disordered pattern.   This inhomogeneity will be an important part of our story. 

In  most  fluids, e.g. water, as the temperature is raised, the difference in density between liquid and vapor gets smaller, until at the critical temperature the difference disappears.    This special point is called a {\em critical point.}   The transition at the critical point is called a {\em continuous phase transition}  because the jump of density between liquid and gas has disappeared.   In the neighborhood of a critical point, a fluid shows a very special behavior.  As one gets closer to the critical  point, fluctuations appear that are almost, but not quite, like the ones found in the boiling region of the first-order transition.   In the boiling region, bubbles of the two different phases show up and can get very large.  Their size is only limited by the size of the container holding the fluid.  Similarly, near the continuous transition, the fluid is an intimate mixture of a huge number of bubbles of vapor and liquid.  The fluid has bubbles of liquid insidw bubbles of vapor, which are themselves inside bubbles of liquid, almost ad infinitum.  The size of the smallest   bubbles is  usually of the order of  the range of intermolecular forces or the distance between molecules. But in contrast to the boiling region, near the critical point  the bubbles' largest size is  limited   by a characteristic length, called the {\em coherence length} that depends upon the pressure and temperature.  This length gives an estimate of the  radius of the largest bubbles.   As one approaches the critical values of temperature and pressure, the coherence length grows, until it is no longer a microscopic distance but is macroscopic in extent.  We can extrapolate this growth process to say that exactly at the critical point, the coherence length will be infinite.  Our concern in this paper will be precisely in understanding the source and consequences of the existence of fluctuation bubbles that have a very large range of sizes.

\subsection{Outline of this paper}
The remainder of this paper is divided into four parts: The next section is concerned with the development and use of the three disciplines that form modern thermal physics: kinetic theory, thermodynamics, and statistical mechanics. Here, we introduce the Ising model and mean field theory as descriptors of phase transitions. 

Section three  describes the early development of renormalization theory.  It is particularly concerned with the use of renormalization methods to circumvent the infinities that appear in all field theoretical formulations used in particle physics.    

Section four explains the basis of the renormalization method as used by Wilson and others to provide connections among physical theories.  It includes discussions of the new insights leading to this renormalization method  as well as the insights that were built into this new approach. We particularly notice that renormalization describes each thermodynamic phase as a different fixed point of a renormalization transformation and suggests that each fixed point represents a different physical theory.

 The next and last major  section of this paper describes the immediate impact of this renormalization theory to the  understanding of phase transitions, particle physics, and dynamical systems.

\section{The Triplet: Kinetic Theory, Thermodynamics, Statistical Mechanics}
Much of physics is deterministic.  Given initial velocities and positions, the laws of classical mechanics will give unique values of the coordinates at a later time.  Equally, given an initial wave function, quantum theory determines the value of the wave function later on.   However, in many physical situations such precision is either useless or unavailable.  If we wish to know the force on the wall of a container containing a macroscopic amount of gas, we should expect an answer to be given in terms of averages or of probabilities.  Calculations based upon the specific particles or their wave functions contain too much information to be useful.  The three disciplines of thermodynamics, statistical mechanics, and kinetic theory have been available for more than one hundred and fifty years\cite{BrushKinetic}, giving us-- with increasing accuracy, depth,  and range-- answers to questions about the properties of materials.  

 \se{ordering} discusses the tremendous diversity of materials available to mankind.  This diversity can, in part, be understood by looking at the phase transitions in  which a material changes from one form of organization and behavior to another, having  qualitatively different properties.   Modern renormalization methods have in large measure developed from a study of phase transitions and the different kinds of order they produce.   In this section,  we look into the three branches of statistical science to reach for the sources of knowledge about materials and their phase transitions. 

\subsection{Kinetic theory}
The basis of this theory\cite{BrushKinetic} was a description of how particles move around in dilute gases.  Kinetic theory is distinguished from pure classical mechanics by having some additional content carried by statistical arguments. 
 
This theory reaches beyond thermodynamics and statistical mechanics in that it describes the details of the non-equilibrium behavior of condensed systems.  Thus, kinetic theory arguments are used to calculate the electrical and thermal conductivity of materials, subjects usually considered beyond the reach of equilibrium theories.   In that sense kinetic theory is considerably more basic and ``fundamental'' than either thermodynamics or statistical mechanics.  However, kinetic theory remains tied to the mechanics of particles and waves and, as far as I know, has not been applied to the statistical and dynamical properties of strings, branes, or other actual or putative constituents of the universe.

\subsection{Thermodynamics\la{thermo}}
Thermodynamics describing the gross properties of materials systems using considerations of conservation of energy, entropy production, and the uniqueness of quantum ground states.  These three ideas are the basis of the three fundamental laws of thermodynamics. These laws entered thermodynamics in sequence and at very different times. 

A central feature of thermodynamics is the use of thermodynamic functions like the Gibbs free energy, which is a function of temperature, pressure, and numbers of particles of various types.  The partial derivative of the free energy with respect to these variables gives us in each case the average value of its {\em thermodynamic conjugate}, thus giving us respectively the average energy, the volume, and various chemical potentials in the system. 

In J. Willard Gibbs' exposition of thermodynamics\cite[Volume I, pp. 55-371]{Gibbs-S}, a phase transition is seen as a singular behavior of a thermodynamic function.   For almost all values of its variables the free energy may be differentiated an arbitrarily large number of times with respect to these variables, without producing any anomalous behavior.  However, at phase transitions this differentiability fails.  At a first-order phase transition, a derivative of the free energy becomes discontinuous.  For example, in the liquid-gas phase transition, the particle density (the number of particles per unit volume) has a discontinuous jump as one passes from the liquid phase to the gaseous one.    

 The use of thermodynamic methods in basic science goes back into the Nineteenth Century, including beautiful work by Maxwell and Gibbs. This use has extended forward into the Twentieth and Twenty-First.  With Landau\cite{quasiparticle1,quasiparticle2} being the most notable proponent, thermodynamic arguments  have been applied to smaller and smaller parts of statistical systems.   The basic actors in modern statistical arguments are long-lived excitations called quasiparticles that can be described by using thermodynamic arguments\cite{Schrodinger}.     

The basic objects of particle physics are best understood as relativistic   quasiparticles,  totally like the ones Landau introduced\cite{quasiparticle1,quasiparticle2} to describe the elementary excitations in He$^{3}$. Indeed, the quasiparticles of the carbon-based material called graphene\cite{graphene} show a kind of relativistic invariance, with the relevant speed being the speed of sound in the material. 

Another recent example of thermodynamic arguments is the Hawking-Bekenstein theory\cite{Hawking, Bekenstein}  of black holes and their entropy. A whole new branch of thermodynamics has been developed to describe black holes\cite{Wald}.

\subsection{Statistical mechanics} 

Kinetic theory can be viewed as a rich and varied collection of methods and results based upon the motion of particles and waves through space.   In contrast the basis of thermodynamics is quite sparse and lean. It can be reduced to a few axioms. But thermodynamics is then supplemented by a rich experience with varied statistical systems.  In this same vein, statistical mechanics is, at its base, a small set of cookbook rules for calculating all the equilibrium properties of statistical systems starting from the Hamiltonian function, $\mathcal{H}$,  which describes the energetics of the system.  In the statistical mechanics of classical particles, for example, we can follow Maxwell, Ludwig Boltzmann, and Gibbs and calculate the average properties of any function of the particles' positions, ${\mathbf q},$ and momenta, ${\mathbf p}$, by using a {\em phase space} composed of these variables. We then take  the probability density for  observing the system at any point in phase space to be given by 
$ d\gamma ~ e^{-[\mathcal{H}({\mathbf p}, {\mathbf q})-F]/T}$ where $\mathcal{H}$ is the Hamiltonian and $d\gamma$ is a volume element in the phase space. 
As a consequence, the average of any function of  ${\mathbf q}$ and  ${\mathbf p}$  is 
\be
<X({\mathbf q},  {\mathbf p} )>  = \int  d\gamma ~X({\mathbf q},  {\mathbf p} ) ~ e^{-[\mathcal{H}({\mathbf p}, {\mathbf q})-F]/T}
\la{canon}
\ee       
   Here,   $F$ is a constant  set to give the proper normalization for the probability
\be
e^{-F/T}=\int d\gamma ~ e^{-\mathcal{H}({\mathbf p}, {\mathbf q})/T}
\la{F}
\ee
Boltzmann\cite{LB} and Gibbs\cite[Volume 2, page 69]{Gibbs-S} both emphasized that $F$ had all the properties of the free energy, as derived from thermodynamics.   Modern statistical mechanics uses \eq{F} to define the free energy. 

Statistical mechanics has maintained its structure since its invention.  It has absorbed quantum mechanics and field theory while retaining its  form as a summation over exponentials defined by the Hamiltonian, as  in \eq{canon} and \eq{F}. 

However, over the years, the content of statistical mechanics has changed quite considerably.  In the late Nineteenth and early Twentieth Centuries, statistical mechanics was almost entirely used to describe gas-like systems in which the interactions among the basic constituents  were very weak. In this situation, the Hamiltonian is a sum over terms, each one describing a single particle or a single mode of excitation.  The statistical sums in \eq{F} can then be performed independently for each of the components. Consequently, the necessary sums can be performed without resort to a computer.      This independent-excitation form of statistical mechanics first served to help us understand the theory of gases derived from classical mechanics and kinetic theory.  Afterwards it was used to support the new quantum theory, and, still later,  to back up views of waves and particles as bosons\cite{Schrodinger} or fermions\cite{quasiparticle1,quasiparticle2}.

\subsubsection{Ising model\la{Ising}}
Starting around 1925,  a change occurred: With the work of Ernst Ising\cite{Ising} and Wilhelm Lenz\cite{Lenz}, statistical mechanics began to be used to describe the behavior of many particles at once.  This approach made it possible to obtain a more exact description of phase transitions. 

Ising and Lenz used a representation of the many-particle system that employed a set of variables,  $\sigma_{\mathbf r}$,  sitting on a spatial lattice.  In their original calculation, each of these variables represented one component of an atomic spin, and could take on two possible values, representing the two possible directions for the spin.  At low temperatures, the majority of these spin  components might preferentially point in one of their two possible directions, and thus form a ferromagnet. For Ising's own work, these variables would sit on a line and would have positions $r=aj$, where $j$ is an integer and $a$ is the distance between lattice sites. In later work, the lattice of spins might get more complex, being perhaps a square lattice ${\mathbf r}=a(j,k)$, or perhaps a lattice in three or higher dimensions.    The usual Hamiltonian for these systems has two terms, one related to a coupling to an external magnetic field and the other reflecting couplings of nearest neighbors.    For example, in two dimensions, the Ising model Hamiltonian takes the form 
\be
-\mathcal{H}/T= K \sum_{j,k}[ \sigma_{j,k} \sigma_{j+1,k}+ \sigma_{j,k} \sigma_{j,k+1}]+ h \sum_{j,k} \sigma_{j,k}
\la{IsingH}
\ee         
where $K$ is a dimensionless number describing the strength of the coupling among spins.  The actual coupling between neighboring spins, with dimensions of an energy, is often called $J$. Then the dimensionless coupling is given by   $K=-J/ T.$  Similarly, $h$ is proportional to the magnetic moment of the given spin times the applied magnetic field, all divided by the temperature.

We shall follow T.D. Lee and C.N. Yang\cite{Lee-Yang,Lee-YangII} and use  the Ising model to describe a fluid.   In this interpretation, $\sigma_r=1$ indicates that the density is higher than average at $\mathbf{ r}$ and $\sigma_r=-1$ indicates a lower-than-average density.  A positive value of $K$ produces an attraction between neighboring regions of similar density, as in a real fluid like water or carbon dioxide.   This interpretation can, with high accuracy,  describe a fluid near the critical point because, as shown by later investigations,  the lattice is quite irrelevant for the description of critical behavior\footnote{The descriptive terms that go with this irrelevance of the lattice are {\em universality} and {\em continuum limit}(see \se{Universality} below) }.  The grainy behavior of a lattice system is replaced by the smooth behavior of a continuum system.  Some of the properties of the system continue to indicate the presence of a lattice, but all the large-scale, long-distance behaviors show a continuum character.  Both the presence of a lattice and also the limited range of variation of the density (originally spin) variable simplify further calculations, making it possible for first Ising and then Lars Onsager\cite{Onsager} to do the sums over spin variables and get exact formulas for the free energy in, respectively, one and two dimensions.  

Far from the critical point, the Ising model's discretized  description of space can only provide a qualitatively accurate description of fluid systems.  However, many useful insights can be obtained from this qualitative accuracy.    

For the Ising model, the statistical mechanics formulae are the same as \eq{canon} and \eq{F}, except that the integral over phase space is replaced by a sum over spins. 

There is a direct interpretation of the parameters in the Hamiltonian. For positive values of $K$ the interaction causes like regions of fluid to tend to sit next to one another, high density with high and lower density with low.  The other parameter, $h$, is  is proportional to pressure.  Higher values of $h$ cause the density to increase and are thus connected with the liquid state while lower values tend toward low densities and gas-like configurations.

\subsubsection{Fluctuating variables in Ising model \la{fluc}}
In statistical mechanics, fluctuating variables that depend upon position are called {\em fields}.  
The field of first importance in the Ising lattice gas  is the dimensionless density of particles:
\be
\rho(j,k) =[\sigma_{j,k}+1]/2
\la{density}
\ee
On part of the line in which the pressure variable, h, is zero the density variable, $\rho(j,k)$,  can take on two values, symmetrically placed about 0.5. (See \fig{mag2}.) The smaller density is descriptive of the gaseous state; the larger the liquid state.  At large positive values of $K$ the two values are close to $\pm 1$; at smaller values they both approach one half. The density stays at one half for values of the coupling, $K$, bigger than the critical value, $K_c$. You can see this from the two curves on the left hand side of \fig{mag2}, which join together and become a single curve on the right hand side of that figure.  

\begin{figure}[t]
\includegraphics[height=8cm ]{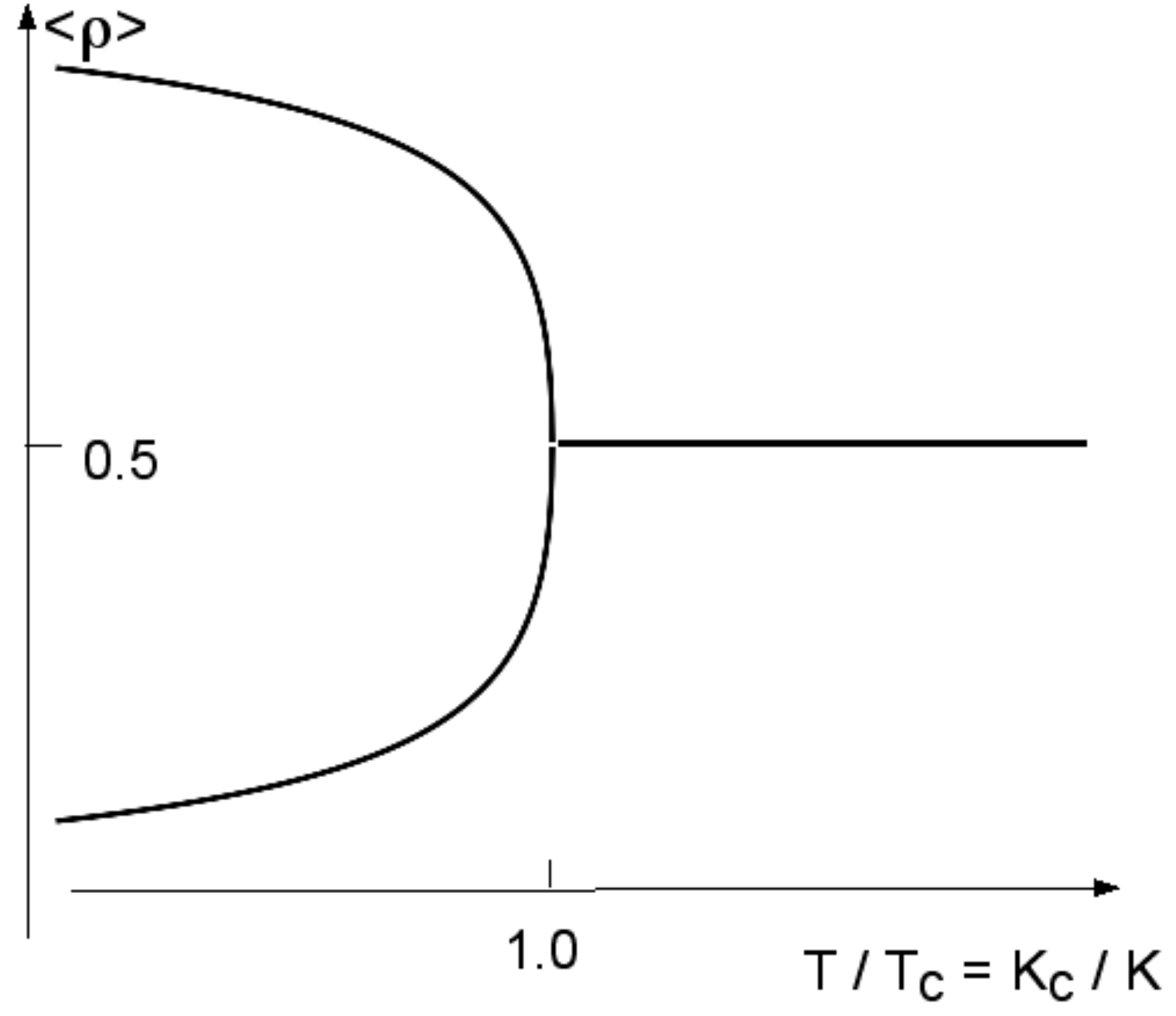}
\caption{Density variable plotted against coupling strength, $K$, or temperature, $T$, for the lattice gas fluid at the critical value of the ``pressure'' field, $h=h_c=0$.  The subscripts ``c'' refer to critical values. }
\la{mag2}
%\end{multicols}%{2}
\end{figure}

Just to the left of the critical point, the difference between the two phases approaches zero.  In his analysis of the Andrews\cite{Andrews} experiment, van der Waals saw that the two density values approached one another so that their difference is
\be   
\rho_{\text{liquid}}-\rho_{\text{gas}}= \text{constant}~\times~ (K_c-K)^\b
\la{beta} 
\ee
where $K_c$ is the value of  $K$ at the critical point.
The critical index, $\b$, has different values in different situations.  For example, $\b$ equals one eighth in the two-dimensional Ising model and is about one third for three-dimensional fluids. In contrast,  $\b$ has  the value one half in van der Waals' mean field theory (see \se{MFT} below).

Another example of a fluctuating field to be found in the Ising model is the energy density
\be
\mathcal{E}(j,k)= \sigma_{j,k} [ \sigma_{j+1,k}+\sigma_{j,k+1}]
\la{energy}
\ee

\subsubsection{Structure of statistical mechanics\la{structure}}
Statistical mechanics has a particularly interesting structure in which the basic fluctuating variables, for example $\sigma_\r $,  serve both as observables and equally determine the  averaging process by being part of the Hamiltonian.  In this section, I describe this structure by considering a generic fluctuating variable at point $\r$, denoted  $o_\a(\r)$.  Here $\a$, as before, describes which variable is under examination. For example, $o_\a(\r)$ might be $\sigma_{\r}$  or it might be the energy density $\mathcal{E}(\r)$.  
     
Now to describe  the structure of statistical mechanics: 

One can envision a quantity like $o_\a$ appearing in the statistical mechanical Hamiltonian.   In particular we may add a term like $h_\a(\r) o_\a(\r)$ to the Hamiltonian, $h_\a(r)$ being a number and $o_\a(r)$ a fluctuating function of basic field variables. The change is:
\bsubs
\be 
- \mathcal{H}/T  \rightarrow   - \mathcal{H}/T  +\sum_r h_\a(\r) ~o_\a({\mathbf r})
\la{peturbr}
\ee
Here, $\r$ refers to a particular spatial position.

In the language of statistical mechanics, variables that appear as products in \eq{peturbr} are said to be {\em conjugate} variables\footnote{This usage is an extension of the definition of ``conjugate'' in thermodynamics as described in \se{thermo}.}. Thus $h_\a(\r)$ and $ o_\a(\r)$ are conjugates.    When the $h_\a$'s are included, the free energy becomes a function of $h_\a$. Because of the structure of statistical mechanics
\be 
- \frac{\partial F/T} {\partial h_\a(\r)} = <o_\a({\mathbf r})>  
\la{partialr}
\ee
where the average is computed with the Hamiltonian of \eq{peturbr} that includes $h_\a(\r)$.    \eq{partialr} gives us a way of calculating the average of $o_\a(\r)$ if we can but compute the free energy.

A further extension of the above argument gives the second derivative of the free energy as another thermodynamic derivative
\be 
- \frac{\partial^2 F/T} {\partial h_\a \partial h_\b} =  \frac{\partial O_\a} {\partial h_\b} =\sum_{r,s} g_{\a,\b}( {\mathbf r,s} ) 
\la{sus}
\ee 
where $g( {\mathbf r,s} ) $ is the correlation function that describes the correlated fluctuations of the local densities $o_\a({\mathbf r})$ and $o_\b({\mathbf s})$
\be 
 g_{\a,\b}( { \r,\s} ) =\big<  o_\a(\r)  o_\b({\mathbf s})      \big>-<o_\a({\mathbf r})> <o_\b({\mathbf s})>
\la{corr}
\ee 
\esubs    
We shall make use of this structure as we describe the history of correlation calculations in \se{OZ}.

\subsubsection {Mean field theory\la{MFT}}

One of the high points of kinetic theory was the development by  van der Waals\cite{Sengers}  of an approximate  equation to describe the relation among the pressure, $p$, volume, $V$, and temperature, $T$, of a fluid.   Soon thereafter Maxwell\cite{Maxwell} improved the theory, allowing it to represent boiling, by inserting thermodynamic stability conditions into the structure of van der Waals' equation. Over the years, statistical mechanics has been used to tweak the van der Waals/Maxwell theory but this theory's content has basically remained unchanged.

Van der Waals' theory basically describes the fluid by calculating its average properties.  Such a description in terms of averages is called a {\em mean field theory}, here abbreviated MFT.  Many different MFTs were derived based upon the thinking of van der Waals and Maxwell.  Each different kind of ordering was  carefully analyzed in terms of its own MFT.  This cascade of work started with Pierre Curie\cite{Curie} and Pierre Weiss\cite{Weiss}, who studied the ordering of magnetic materials like iron.  All these mean field theories have qualitatively similar properties, especially near their critical points. The thermodynamic properties of this kind of theory was extensively discussed in paper I\cite{I}, and its defects, especially near the critical point,  outlined in paper II\cite{LPKwebsiteII}. In this paper, we shall go over some of the same ground, but focus our attention upon the correlations among the behaviors in different regions of a given material.

  The model fluid described by the equations of van der Waals and Maxwell   showed a behavior that was qualitatively similar to the properties of carbon dioxyde, as observed by Andrews\cite{Andrews}.

Mean field theory can describe quite well how a particular molecule can influence molecules in its immediate neighborhood.   Thus, in particular, the van der Waals theory can and does say both that
\begin{itemize}
\item a higher density of molecules in one region of the fluid tends to attract molecules into neighboring regions, and
\item  conversely, a lower density of molecules in some other region will tend to induce neighboring regions to have a lower-than-usual density  
\end{itemize} 
Both of these effects are directly included in the MFT.

Moreover, mean field theory goes much further than just indicated.  It not only includes direct effects:  ``My neighbor helps me line up with him,'' but also long chains of such effects as in a situation in which everyone's average order helps encourage everyone else. The long-range order produced by overlapping and mutually sustaining regions of local order is the basic source of nature's different phases.  These linked orderings are certainly included in MFT calculations.    

What is not included is fluctuations.   If some portion of the system departs substantially  from average behavior, mean field theory cannot describe it.  MFT is positively hopeless in its description of ordering in a system arrayed in a linear fashion in which each component is connected with nearby ones to its left and to its right.  In this case, as pointed out by Landau\cite{1d}, a fluctuation, perhaps a very  exceptional one, can always break the link that connects one part of the system with another.    In a very long system, the breaks will happen many times, ruining the long-range order, and thereby preventing the occurrence of any phase transitions.   This phenomenon is totally outside the range of description of MFT.    In fact, MFT incorrectly predicts phase transitions in one dimension.

In higher dimensions, many paths link two well-separated points. Correlations along these different paths work together to produce ordering.  The higher the dimension, the harder it is for fluctuations to ruin the correlations produced on all these paths. Therefore, as dimension is increased,  MFT
 gets successively more accurate.   In particular, this kind of theory is much better at treating spatially extended order for three-dimensional hunks of matter than for ordering of thin films spread along surfaces.

In two dimensions, MFT often
 gives qualitatively incorrect predictions for phase transitions.  The three simplest two-dimensional models have nearest-neighbor couplings of local vector variables, $\mathbf{S}(\r)$.   In the three cases--- called respectively the Ising model, the XY model, and the Heisenberg model--- the vector $\mathbf{S}$ has respectively one, two, and three components and fixed magnitude. In all three cases, the couplings between neighbors are of the form $K \mathbf{S}(\r) \cdot \mathbf{S}(\s)$. However, the three cases have different outcomes: the Ising case has a finite temperature phase transition and long-range ordering\cite{Onsager,Yang}; the XY situation has a finite temperature phase transition\cite{KT, MerminWagner,Hohenberg, Coleman} and an ordering that decays with distance (see \se{XY} below); the Heisenberg model  has neither phase transition nor extended order.   MFT
 gives all three systems the same behavior:  a phase transition plus long-range order,  and thus gets two out of the three totally wrong.

\section{Field Theory Formulations}
We shall argue that renormalization methods are an outgrowth of phenomenological studies of phase transitions.  But these methods are equally a consequence of the development of expansion methods  in  field theories.
\subsection{Classical electrodynamics}
One of the earliest field theories, due to Hendrik Lorentz\cite{Lorentz}, describes classical electrodynamics, including both Maxwell's equations and the motion of  electrons.  In the Lorentz theory,  classical electrodynamics can be described by elementary processes in which electromagnetic waves are absorbed and emitted and in which particles are deflected by electromagnetic fields. The elementary processes are combined together to describe in a step-by-step fashion the complex events that can result from the field theory. These calculations can be tracked and described by pictures akin to modern Feynman diagrams. 

This simplest field theory describes point-like electrons. Composite steps in this field theory tend to  contain divergent terms most often caused by processes occurring on a very short length scale but also sometimes by processes on a very large scale.  These divergences occur precisely because we use the very same theory to describe the processes that occur at all length scales.  We do this because we are starting from a theory with wonderful symmetry properties  like   relativistic invariance, local interaction, gauge invariance, and causality, and we are unwilling to compromise these  symmetry properties. 

In classical electrodynamics, the most obvious divergences arise because the electric field configuration around a point-like electron gives this region of space an infinite energy.  In a relativistic expression for the theory, the infinite energy implies an infinite mass and then predicts some very strange events.  It is certainly logical to ask whether that kind of electron can be incorporated into a consistent electromagnetic theory\cite{Pietsch}.  The struggle to do this started with Lorentz\cite{Lorentz}, who tried to get around the problem by introducing an electron with a non-zero radius  and continued with Paul Dirac\cite{Dirac},   who used a point electron, but included a subtraction term to make the electron-mass finite.       Their approaches both  disagree with causality and relativity, and were thereby ruled out.  Later on Dirac\cite{DiracBS} made use of an approach more like Lorentz's but employing some quantum theory, using  Bohr-Sommerfeld semi-classical quantization to generate an excitation with a mass approaching that of a muon. I do not think Dirac's approach has been extensively followed up. 

\subsection{Renormalization  in quantum field theory}  
   
In the meantime, quantum electrodynamics had been developed. The earliest quantum renormalization study was apparently that of Hendrik Kramers\cite{Kramers}.   Physics began to see renormalization as a pressing need just after World War II when the Dirac equation failed to explain a splitting of two energy levels in the hydrogen atom.  The Dirac theory made these two energies identical; Willis Lamb and Robert Retherford\cite{Lamb} measured the difference between the levels and found a non-zero splitting.

Like the classical theory, the quantum field theory predicted an infinite electron mass\cite{Schweber}. This infinity was potentially a roadblock that would prevent the most direct ways of predicting the magnitude of the splitting.   An approach  for  going around this infinity was suggested by Kramers\cite{DysonPT}, and was followed by Hans Bethe in a non-relativistic calculation. Then  Richard Feynman, Sin-Itiro Tomonaga, and Julian Schwinger did a fuller and more relativistic calculation. These three developed methods that could be  applied to understand the Lamb shift and kindred effects in the quantum field theory of electrodynamics.   Their {\em renormalization} methods have a weak mathematical foundation because of the infinities inherent in the field theory, but nonetheless they give answers to problems in electrodynamics that agreed in excellent fashion with experiment.   

Renormalization's expansion methods work well in electrodynamics because of this theory's weak interaction between electrons and radiation. The strength of the interaction is measured by a dimensionless constant, $\a \approx 1/137 $.  
Similar renormalizations were required for other field theories, but they did not initially give accurate results because the other theories did not have an equally useful small parameter.    

The basic ideas supported by renormalization methods were described in the context of field theory by E. C. G. Stueckelberg and A. Peterman\cite{SP}. They pointed out that renormalization calculations depended upon dividing the Lagrangian density defining the theory into two parts: a ``bare'' density and interaction terms.  The former serves as a starting point while the latter is used to generate a step-by-step expansion.   These authors pointed out that the end-result should not depend upon the particular division.  Since the result was independent of a group of transformations in the formulation of the problem, Stueckelberg and Peterman described the approach as the ``renormalization group''.  Later calculations were less focused upon the expansion techniques.  Nonetheless, the name stuck.  Murray Gell-Mann and Francis  Low \cite{GML} then pointed out that the renormalization method might be used to describe how the observed properties of quantum electrodynamics, as seen at a particular length scale, might then depend upon the length scale of the observations.  This insight proved crucial for later work. 

\subsection{Renormalization is ugly}
Renormalization ideas were often considered somewhat suspect and perhaps even ``ugly,''\footnote{Freeman Dyson quoting Dirac on renormalization: ``I might have thought that the new ideas were correct if they had not been so ugly.'' } perhaps because they were on a very insecure mathematical footing.  Another reason to be dissatisfied with renormalization is that it does not fit very well into the {\em ethos} of particle physics. Particle physics has always tried to construct a theoretical approach that could be ``fundamental'' and apply always and everywhere.     A process of renormalization is necessary precisely because the theory at hand is marred by infinities. For that reason the theory is mathematical nonsense\footnote{One reason that many physicists have been attracted to string theory is precisely that this theory has no infinities and thus requires no renormalization. }.  When the nonsense is swept under the rug by a renormalization procedure, the result does not fit with all our preconceptions about the nature of a ``final theory'' that might well describe everything without internal contradiction. In fact, when carefully examined the renormalized  theory usually fails\cite{Fraser} to obey causality or relativistic invariance or to show a really simple structure, or maybe it simply fails to be predictive.   Usually the failures occur because an extrapolation, often to a different regime of length scales, makes the theory look foolish.  This kind of situation is not distressing in, say, condensed-matter physics where we expect each of our theoretical models to be only an approximation, and we know that the models can only be pushed so far.  However, in particle physics one is tempted to demand  that a proper theory can have no limits in its range of validity.         

\subsection{Quantum field theory=statistical mechanics}
Before his work on quantum electrodynamics, Richard Feynman\cite{path} was responsible for a remarkable reformulation of quantum theory in terms of path integrals.  Even more remarkably, this reformulation permits a simple and direct mapping that enables any quantum problem in $d$- dimensions to be expressed in terms of a statistical mechanics problem in $d+1$ dimensions\cite{LPK00} using analytic continuation (\se{continuation}).  Conversely any lattice statistical mechanics problem in $d+1$ dimensions that has basic variables that are coupled via nearest-neighbor interactions in one of the dimensions is immediately convertible into a quantum mechanics problem in $d$ dimensions. 

Kenneth G. Wilson strongly emphasized this connection between quantum theory and statistical mechanics.  He used it to phrase problems in quantum field theory in terms of equivalent or almost equivalent problems in statistical mechanics.   In particular he pointed out that the formulation of phase transition problems using the Landau form of the free energy (see paper I\cite[page 21]{I}), with Vitaly Ginzburg's inclusion of fluctuations\cite{Ginzburg} could  equally well be used to define problems in quantum field theory and phase transition theory.  Calculations thus formulated were then described as using the Landau-Ginzburg-Wilson action or free energy.  The major difference between particle physics and condensed-matter calculations was that the former were usually carried out at four dimensions whereas the latter were most often done at lower dimensions.  

 I  describe here how to set up the statistical physics problem most simply connected with quantum field theory.  Let me imagine a problem arranged on a four-dimensional simple cubic lattice with a coordinate of the form $x=k,l,m,n$, with $\r=l,m,n$ being a ``space'' coordinate and $k$ a coordinate connected with time development.  In what follows, we shall make ``space'' and ``time'' perfectly symmetrical.  Our basic variable will be written as $\phi(x)=\phi(k,\r)$.  Here $\phi$ is real and runs from minus infinity to infinity.   Our problem will have a ``Lagrangian density'' 
\be
\mathcal{L}(x) = -\sum_{y ~n.n.~ to~ x} \frac{[\phi(x)-\phi(y)]^2}{2}+h(x)~\phi(x)- m_0^2 \phi(x)^2-\lambda ~ \phi(x)^4
\la{lagrangian} 
\ee
Here the sum is over all $y$'s that are nearest neighbors to $x$.     
This first term produces correlations among neighbors; the second term gives a handle or {\em source}, $h(x)$, that can be used to calculate averages;  the third term  provides a mass limiting the range of correlations; and the last provides an interaction describing the scattering of particle-like excitations. Using this Lagrangian density, one calculates a free energy in a manner very similar to the one we used for the Ising model. First one calculates the total Lagrangian by summing the Lagrangian density over the entire lattice.  Then, the free energy is formed by integrating the exponential of the Lagrangian over all the statistical variables.  In symbols,
\bsubs
\be
e^{-F\{h\}/T} =\Big[\prod_x \int d \phi(x) \Big] ~~\exp[~\sum_x \mathcal{L}(x)]
\la{action}
\ee
In this expression, $F/T$, called an action in the context of field theory,   is a function of the applied field or {\em source}, $h$, defined at every space-time point\footnote{The notation$\{h\}$ indicates that the action is a function of the value of the entire source function, not of the value at any particular point. }.

\eq{action} for the action generates averages in just about the same way as does \eq{F} for the free energy.  In direct analogy to \eq{partialr} the ground state field average of the quantum field is given by 
\be
< \phi(\x)  > =-\frac{\d F/T\{h\}}{\d h(x)} 
\la{partialQ}
\ee

In direct correspondence to \eq{corr}, which defines the statistical correlation function, one can generate an expression for the correlation between the fields at two points by 
\be 
 G ( x,y ) =< \phi( x)  \phi( y)    >- <\phi(x) > <\phi(y) > =-\frac{\d^2F/T}{\d h(x) ~\d h(y)}
\la{corrQ}
\ee  
\esubs
When this correlation function appears in quantum field theory it is called a Schwinger function.

\section{Phase Transitions and Renormalization}
\subsection{Extended singularity theorem}
The crucial fact in the study of phase transitions is that, as one passes through such a  transition,  there is an abrupt change in form for the thermodynamic free-energy function, $F$.    Different phases have different behaviors and the transition from one behavior to another engenders an abrupt change in $F$.  In mathematics, an abrupt change is called a singularity and is usually signaled by an infinity in a derivative of the function in question.  Singularities are not possible in a statistical mechanical result for a finite system.  The integrals in \eq{canon} and \eq{F} are basically just sums of exponentials formed from  $\exp(-\mathcal{H}/T)$.  The Hamiltonian itself is a smooth function of temperature and pressure for the Ising model.   In the expression of \eq{canon} the inverse temperature appears as a multiplicative parameter and the volume in the limit of integrations over the $q$'s. Thus this equation implies that the free energy is a sum of smooth functions of temperature and volume.  Such a sum is itself a smooth function.    We conclude that, in any finite system, the free energy contains neither singularities nor abrupt changes.

To have a singularity, it is necessary that the free energy contain the result of summing over an infinite system.  (In a previous publication, \cite{LPKwebsiteII},  I gave a name to this old result and called it the {\em extended singularity theorem}.)   So, for a sharp phase transition, we must have a statistical mechanical system that is not finite but infinite in extent with an infinity of different summation or integration variables.   Because of the structure of statistical mechanics, the extended singularity theorem has a number of important consequences, particularly near a critical point, a place where thermodynamic derivatives diverge. 
\subsubsection{From critical opalescence to Ornstein and Zernike\la{OZ}} 
Observers have long noticed that, as they move close to the liquid-gas critical point, the fluid, hitherto clear and transparent, turns milky.  This phenomenon, called {\em critical opalescence}, was studied by   Marian Smoluchowski (1908)  and  Albert  Einstein (1910)
\cite[p. 100]{Pais}.  
Both recognized that critical opalescence was caused by the scattering of light from fluctuations in the fluid's density.  They pointed out that the total amount of light scattering was proportional to the compressibility, the derivative of the density with respect to pressure\footnote{As pointed out to me by Hans van Leeuwen, the opalescence is very considerably enhanced by the difficulty of bringing the near-critical system to equilibrium. This out-of-equilibrium system tends to have anomalously large droplets analogous to those produced by boiling. These droplets then produce the observed turbidity. So the infinite compressibility that occurs at the critical point is only a portion of the explanation of the opalescence effect. }.  They also noted that the large amount of scattering near the critical point was indicative of anomalously large fluctuations in that region of parameters.    In this way, they provided a substantial explanation of critical opalescence\footnote{Einstein then used the explanation of this physical effect to provide one of his several suggested ways of measuring Avogadro's number, the number of molecules in a mole of material.}.  

But this compressibility explanation was quite incomplete.  What causes the divergence in the compressibility?  The structural analysis of \se{structure} points the way to an answer to this question.  According to statistical mechanics, the thermodynamic derivative studied by Smoluchowski   and   Einstein, $ {\partial \rho({\mathbf r})}  / {\partial p}$, is  given by the correlation function expression
\be
\frac  {\partial \rho({\mathbf r})}  {\partial p}= \sum_s g(\mathbf {r,s})=\sum_s\big[
\big<[\rho({\mathbf r})-<\rho({\mathbf r})>]~[ \rho({\mathbf s})-< \rho({\mathbf s}) >\big >\big]
\la{CorrToThermo}
\ee      
The next question is what happens to make the sum infinite: Do individual terms in the sum diverge?  

This question was soon answered.  Leonard Ornstein and Frederik Zernike\cite{OZ} used kinetic theory and the approach of van der Waals to derive an approximate form for the density correlation function, $g$, in the form
\bea
 g(\mathbf {r-s})&=& \text{constant} \times\int \frac {d^d \mathbf k}{(2\pi)^d}
\frac{\exp i \mathbf {(r-s)\cdot k }/a}{ k^2+ |T-T_c|/T_c}     \nonumber   \\
&=&\text{constant} \times(a/|{\mathbf {r-s}}|) \exp( - |{\mathbf {r-s}}|/\xi)
 \la{OZeq} 
\eea 
The first line applies to all dimensions, $d$; the second line holds for a system in three dimensions. Here $a$ is, once more, the distance between adjacent lattice sites.  The physics of this  result was that the scattering is produced by small regions, droplets, of materials of the two different phases in the near-critical fluid.  The regions would become bigger as the critical point was approached, with the droplets extending over a spatial distance called the correlation or coherence length, $\xi$.    Ornstein and Zernike saw this length diverge on the line of coexistence between the two phases of liquid-gas phase transition as the critical point was approached in the form
\be
\xi =  a (T_c/|T-T_c|)^\nu     \text{~~with~~}  \nu=1/2
\la{xi}
\ee
with $T-T_c$ being
the temperature deviation from criticality\footnote{The use of the symbol, $\nu$, in \eq{xi} is totally anachronistic.  Long after Ornstein and Zernike, scientists from the King's College School (Cyril Domb, Martin Sykes, Michael Fisher, ...) set a standard notation\cite{Domb} for the various exponents used to describe critical phenomena.  The index corresponding to the coherence length was set to be $\nu$.}.    Thus the correlation length goes to infinity as the critical point is approached.   This divergence is crucially important to the overall understanding of the critical point. 

The Ornstein-Zernike result, \eq{OZeq}, only applies when the separation  distance,  $|{\mathbf{r-s}}|$, is large in comparison to the distance between lattice points, $a$.  Therefore the predicted correlation function never grows very large.  The compressibility, arising from the spatial sum of the correlation function, does get large as the critical point is approached precisely because the correlation length, $\xi$, diverges.  Therefore the sum includes larger and larger droplets as one gets closer to criticality.  The factor of  $1/|{\mathbf {r-s}}|$ gets smaller but the summation nonetheless grows as $|{\mathbf{r-s }}|^2$ in three-dimensional space.  Precisely this growth is responsible for the divergence in the susceptibility and the corresponding divergence in the light scattering.  

The analogy between quantum calculations and statistical ones includes the behavior of correlation functions.  The Ornstein-Zernike form of the correlation function ( \eq{OZeq} ), originally used in condensed-matter physics, reappears in the simplest version of the Schwinger function (\eq{corrQ}) used in field theory\cite{Yukawa}. This function is exactly the same in the two theories, except that the parameters are different in the two theories.     Where we have the squared mass, $m^2$, in \eq{corrQ}, the approximate condensed-matter physics result uses $(T-T_c)/T_c$.    Approach to the critical region in the statistical theory corresponds to the small-mass limit in the field theory.

\subsubsection{The ordering } 
Many of the qualitative properties of the critical point can be seen in the behavior of the average density $ <\rho(\r)> $ (as described in \se{fluc}) and of the density correlation function of \eq{corr}.

The criticality is reflected  both in the exact correlation functions and also  in the approximate result calculated by Ornstein and Zernike (see  \eq{OZeq}).   Away from the critical point, the sum in \eq{CorrToThermo} converges because, when $\mathbf r$ and $\mathbf s$ are separated by an amount that is large compared to the coherence length the average of terms like $\rho(\r)-< \rho(\r)>$ remains small.  However, as $\r$ comes closer to $\s$ the observed fluctuations become larger  because the system does not ``know'' whether it is a liquid or a gas and it fluctuates back and forth between densities appropriate for the two phases.  The two possible values describe respectively the liquid and the gaseous situation.   The fact that, under identical conditions, the system can take on one of several (here two) possible values is described by saying the system has {\em long-range ordering} or simply that it has taken on {\em order}.  It is interesting and somewhat surprising to notice that a simple fluid system can, under identical conditions, fall into one of two possible states (liquid or gas) and that this state can persist over arbitrarily long distances.     

The behavior calculated by Ornstein and Zernike is not quantitatively correct because the van der Waals theory is not perfect, but the general direction is quite right.  Near the critical point, a variety of different thermodynamic derivatives diverge because the correlation functions fall off as some power of distance while the correlation length diverges.   This is the qualitative structure exhibited by the Ornstein-Zernike theory and is equally the qualitative structure exhibited by the exact theory.

In one respect the liquid-gas phase transition is atypical.  Most phase transitions produce a breaking of some symmetry principle.   For example, in the Heisenberg model of a ferromagnet the basic variables are spins that can point in any direction in three-dimensional space.  The disordered state that exists at  high temperatures has full rotational symmetry in which all directions are equivalent.  In the ordered state that arises at lower temperatures,  the spins gain an average direction so that they preferentially point in some one of the spatial directions.  The rotational symmetry has been lost in the phase transition.   The critical point, in between the two regions, retains the fully rotational symmetry.  However, local regions of the system will tend to show spins that align in some particular direction.   They serve as a sort of {\em fluctuation droplet} of a particular ordering.  
\subsection{The renormalization method: Bridges among theories}
Maxwell\cite{Maxwell-Gases}, Boltzmann\cite[Giovanni Gallavotti, page 54]{LB}, and Einstein\cite[page 100]{Pais} all  provided bridges between the mechanistic and atomistic view of kinetic theory and the more axiomatic subjects of thermodynamics and statistical mechanics. In the Twentieth Century, condensed-matter physics provided similar links between a molecular view, often based upon quantum theory, and more phenomenological descriptions of materials--- based upon elasticity theory, acoustics,  and materials engineering.   I focus upon one important gap in these theories that existed through the 1970s:   They were weak in describing situations, as for example, the ones near phase transitions, in which  variations in material properties on a large spatial scale  play a substantial role.   The extended singularity theorem insists that this kind of  fluctuation is important for determining how different phases of matter fit together. 
   
This hiatus in theoretical knowledge was closed by the development of  renormalization techniques\footnote{This method goes under the somewhat misleading name, {\em renormalization group}, because the primitive version of the theory had the mathematical structure of a group, in contrast to its more developed form, which is a semi-group.}. This method was put together after a host of theoretical and experimental studies gave clear indications that mean field theories were wrong\cite{Domb}.  In paper II, see Ref. \cite{LPKwebsiteII}, I described some of the theoretical and experimental evidence pointing to the failure of mean field theory to give an accurate picture of behavior near the critical point.   In \se{MFT} I gave additional evidence indicating that in low dimensional systems MFT often failed to predict correctly whether or not  a phase transition could occur.  Clearly, some additional theory was needed.

\subsection{The renormalization method in statistical physics}
Statistical mechanics describes the equilibrium behavior of thermodynamic systems in terms of a probability distribution, termed an {\em ensemble}, that describes the probability of occurrence of all the possible configurations of the system. The equilibrium nature of the ensemble is built into the requirement that the ensemble not change as the particles move about, obeying the rules of classical mechanics. Boltzmann showed how to construct such ensembles given a knowledge of the energy of each configuration\cite{Uffink}.   The basic problem faced by workers in statistical mechanics is to compute the observable properties of a system given a specification of its ensemble, usually given in terms of such basic parameters as temperature, chemical potential, and the form of the interaction among particles.  

Direct calculation of ensemble properties has proven to be very difficult for systems at or near phase transitions.  Here the behavior is dominated by correlated fluctuations extending over regions containing very many particles. As the phase transition is approached, the fluctuations cover larger and larger regions, over distances finally tending to become infinitely large.   Because of this infinity, computing the correlations among the different particles presents a calculational problem of very considerable complexity.   This problem was unsolved before the development of the renormalization method.

This method, brought to its completed form by Wilson\cite{Wilson}, provides an indirect method of approaching the difficulty presented by large fluctuating regions.  The renormalization method, a process of averaging the ensemble  over small-scale correlations, constructs a new ensemble that describes the same problem, but uses spatially averaged variables to do so.  Step by step, the gaze of the investigator focuses upon larger and larger length scales. It is usually true that eventually, after many renormalizations, the system will reach a  {\em fixed point}, that is, an ensemble that remains unchanged under the successive transformations.      Methods have been developed that enable one to calculate the new ensemble, using the larger length scale, from the old one and thereby find the ensembles describing the fixed points.  These calculations are the bread and butter of renormalization theorists.

\subsection{Block spin transformation\la{Block}}
I describe here a basic method of formulating a renormalization calculation\cite{LPK1966}.  We start from a free energy calculation as in \eq{F}
\be
\exp({-F_{\{K\}}/T})=\sum_ {\{c\}} ~ \exp({-\mathcal{H}_{\{K\}}{\{c\}}/T})
\la{Fc}
\ee  
expressed in a form appropriate for discrete variables like those of the Ising model. Here, the appearance of the  $\{K\}$ indicates that $\mathcal{H}$ and $F$ depend upon a group of coupling constants like $K, h, \cdots $.  The $\{c\}$ denote the possible configurations of the system, that is, the values of all the statistical variables, perhaps the $\sigma$'s or $\phi$'s mentioned above.    To do the summation in \eq{Fc} we  imagine redoing the sum using a new set of variables constructed by splitting the system into cells on a new rougher lattice described by the coordinates $\R$ and the new spin variable $\mu_R$.   Each new spin variable is intended to summarize the situation in a block containing several old spin variables. (See \fig{block}.) To make that happen one can, for example,   pick the new variables to have the same direction as the sum of the old spin variables in the block and the same magnitude as each of the old variables. The change could then be represented by saying that the distance between nearest neighboring lattice sites would change from its old value, $a$, to a new and larger  value, \bsubs \la{rg}
\be
a' =\ell a.
\la{ell}
\ee
See \fig{block}, in which the lattice constant has grown by a factor $\ell=3$.

We would then sum over the old spin variables, holding the new ones fixed.  This kind of change can be represented in symbols by writing the new summation as
\be
e^{-\mathcal{H'}     
\{C\}/T   }   =\sum_ { \{c\} } 
 \mathcal{B} (\{C\},\{c\}) ~
 e^{-\mathcal{H}{\{c\}}/T}.
\la{Hrenorm}
\ee
Here $\mathcal{B} $ is a function of the new spins and also the old ones.  This function defines the blocking method. It is required to obey the condition that the sum over new spins obeys
\be
\sum_ { \{C\} } 
 \mathcal{B} (\{C\},\{c\})   =1
\la{Crenorm}
\ee
This condition ensures that the new free energy calculated from the new Hamiltonian, $\mathcal{H'}$ has exactly the same value as the old one. Hence the renormalization procedure holds the free energy constant while changing the lattice as in \fig{block} and changing the variables on the lattice from the $\sigma$'s to the $\mu$'s.        
\begin{figure}
%\begin{centering}
\begin{multicols}{2}
\includegraphics[height=8cm ]{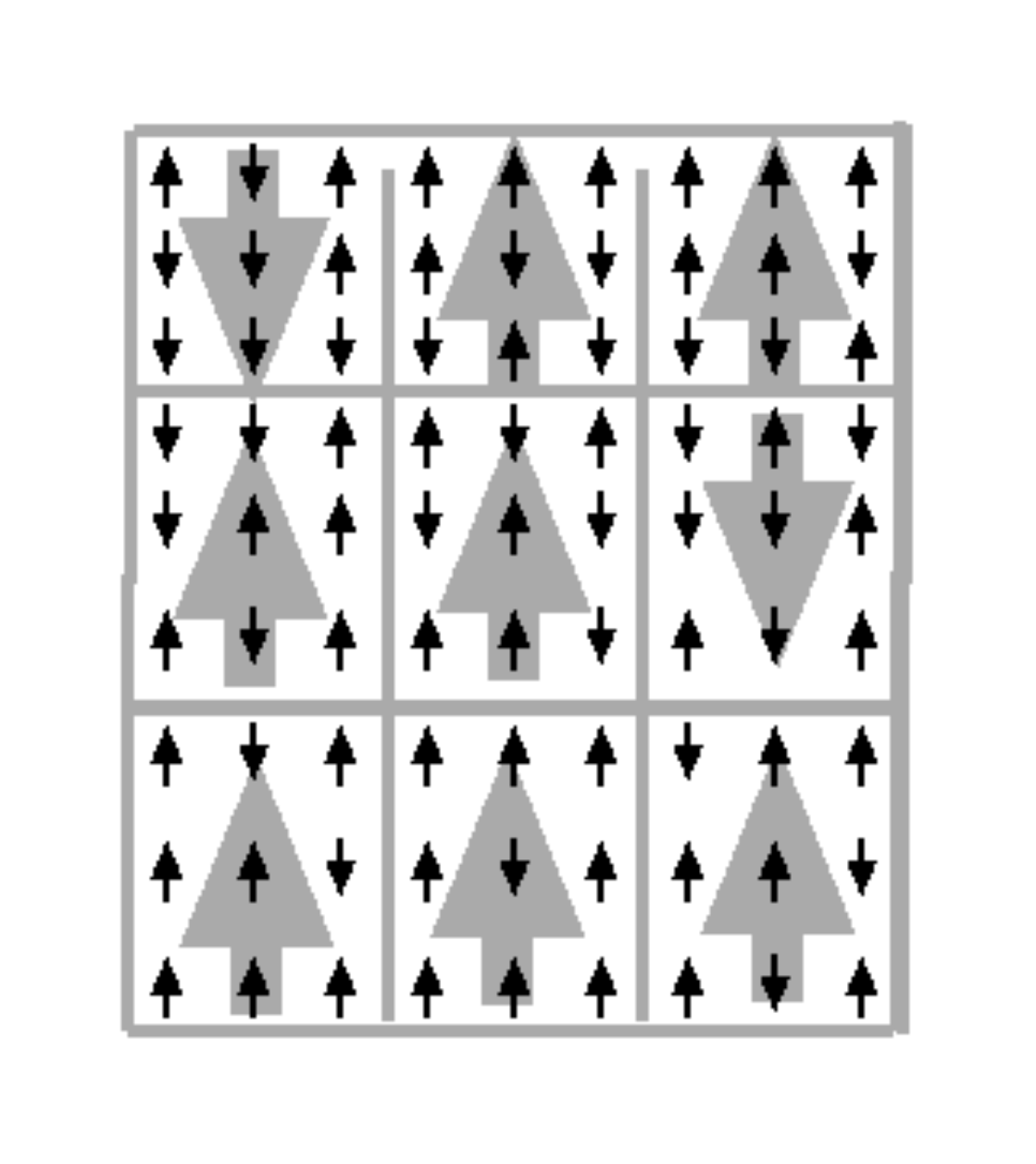}
\caption{ Making blocks. In this illustration a two-dimensional Ising model containing 81 spins is broken into blocks, each containing 9 spins.  Each one of those  blocks is assigned a new  spin  with a direction set by the average of the old ones. We imagine the model is reanalyzed in terms of the new spin variables.  }
\la{block}
\end{multicols}%{2}
\end{figure}  
 
  One can then imagine doing an approximate calculation to set up a new ``effective''  free energy calculation that will give the same answer as the old calculation based upon an approximate ``effective'' Hamiltonian making use of the new variables. 

\subsubsection{The Wilson insights}
In 1971 Kenneth G. Wilson was well aware of both the statistical mechanics and the field theory work that we have outlined here.  He had been a student of Murray Gell-Mann, who had been coauthor of one of the seminal papers on the renormalization method in particle physics\cite{GML}.   In this early paper, renormalization was applied not once but many times so that one could produce big changes via small steps. Wilson had also been studying the statistical-mechanical approaches for many years, so that he had full control of the material outlined above.  He was ready to produce a major advance.
   
Previous workers set up the effective Hamiltonian by using some subset of the conceivable couplings in the system, as in \eq{IsingH}. This left one with a somewhat uncontrolled approximation scheme that was hard to use. 

One essential and brilliant insight that Wilson added to the calculation is to ask what would happen if one would include all the possible couplings consistent with translational invariance and the other symmetries of the problem.  Then one could write the Hamiltonian on the right-hand side of \eq{Hrenorm} as
$
\mathcal{H}_{\{K\}}{\{c\}}
$ 
where $\{K\}$ is a list all the possible couplings.  The renormalized Hamiltonian, appearing on the left-hand side of that equation would have exactly the same form as before, but depend upon the new spins, and be parametrized by new values of the couplings, $K', h', \cdots $.  We could therefore write the new Hamiltonian as
$
\mathcal{H}_{\{K'\}}{\{C\}}
$ 
The  goal is to find the function $\Re$ that gives the new couplings as functions of the old ones:
\be
{\{K'\}}=\Re({\{K\}})
\la{renorm}
\ee
\esubs
\eqs{rg} describe the mathematical structure of the renormalization group.

\subsubsection{Approach to fixed point\la{fixed}}
The next major insight of Wilson was to ask what would happen if one did many successive renormalization calculations. Eventually, one of three things could happen:
\begin{itemize}
\item One possibility occurs away from the phase transition and in smaller systems.  After some number of transformations the scale produced by the renormalization exceeds the size of the largest correlated regions.  At that point, the couplings among the renormalized particles or lattice sites become weak and we reach a fixed point descriptive of uncorrelated excitations. Given the weak couplings, the properties of the renormalized ensemble can be evaluated, and the properties of the original ensemble inferred from that result.  
\item The second possibility is that the ensemble reached by many renormalizations may include strong couplings and infinitely long-range correlations.  In this case, the result is indicative of a first-order phase transition.  The strong couplings provide another kind of fixed point that enables one to evaluate the behavior at the first-order transition. 
\item The third possibility occurs at the critical point, where the successive renormalizations eventually give a nontrivial fixed point, that is an ensemble with finite, non-zero and non-infinite, couplings that remain fixed under the transformations.  The properties of this ensemble enable one to infer the main aspects of behavior at or near the critical transition.
\end{itemize}

\subsection{The critical fixed point}
The qualitative properties of the critical fixed point provide an important legacy of the renormalization approach.  Some of these will be described in this section.  However, the philosopher of science and the historian of science will be equally interested in noting that while the renormalization group method gave the substantial theoretical underpinning for the material of this section, much of the phenomenological content of the section was known before 1971.
\subsubsection{Classification of couplings\la{relevant}}
Since the Wilsonian point of view generated the renormalization of many different couplings, it became important to keep track of the different ways in which the couplings in the free energy would change as the length scale changes.  This work starts with an eigenvalue analysis. Let the values of the couplings at the fixed point be written as $K^*_\a$, where as before $\a$ distinguishes the different couplings.  Denote  the deviation of the couplings  from the fixed point value by 
\be
h_\a= K_\a-K^*_\a
\la{deviation}
\ee
After the renormalization, the new deviations obey
$$
h'_\a= K'_\a-K^*_\a
$$
Now comes a natural but crucial step.  One takes linear combinations of couplings and arranges the combinations so that, after a renormalization, every combination reproduces itself except for a multiplicative factor.   In other words, this approach makes every linear combination of couplings obey the equation\cite[chapter 12]{LPK00} 
\be
h'_\a=\ell^{y_\a} h_\a
\la{ydef} 
\ee    
The different combinations are then classified according to the values of the index, $y_\a$, which may be complex.   There are  three possibilities\cite{DGL}[see F. Wegner, Vol. 6,1976, p.8]: 
\begin{itemize}
\item   Relevant:  Real part of $y_a$ greater than zero.   These are the couplings like $t$ and $h$ that grow larger as the length scale is increased.   Each of these couplings will, as is grows, push the material away from the critical point.  In order to reach the critical point, one must adjust the initial Hamiltonian so that these quantities are zero.
\item Irrelevant: Real part of $y_a$ less than zero.   These couplings will get smaller and smaller as the length scale is increased so that, as one reaches the largest length scales, they will have effectively disappeared
\item Marginal: Real part of $y_a$ equal to zero.   
\end{itemize} 
The last case is rare.  Let us put it aside for a moment and argue as if only the first two existed.

 \subsubsection{Additional scalings}
 We describe \eq{ydef} by saying that $h_\a$ scales as $\ell^{y_\a}$, or equivalently by saying that $h_\a$ has scaling index $y_\a$.   The content of this statement is that as the length scale is changed the coupling represented by $h_\a$ grows in the same way as $\ell^{y_\a}$.  Of course, this ``growth'' is really a decay when the real part of $ y_\a$ is negative.    
 
These distictions, albeit in different words, can be found in the work of Benjamin Widom\cite{Widom,WidomII,LPK1966}.  Note that the exponential of the free energy  defines a probability, so that the free energy can be said to have scaling index zero.   Then, by virtue of \eq{peturbr} one can say that the conjugate operator $o_\a$ has index $d-y_\a$.   Since this density is given the index, $x_\a$, we derive the relation
 \be
x_\a=d-y_\a 
 \la{hyper}
 \ee
 In this way, we can ascribe scaling indices to the local operators that appear in the theory.  
 
There is a simple way of characterizing the scaling properties of the various quantities that appear in the critical-point theory. While  $\r$ changes by a factor of $\ell$ under a renormalization transform, $\r \rightarrow \ell ~\r$ we say that $o_\a( \r)$ scales like $r^{-x_\a}$.   This means most specifically that at the critical point all correlations of the local operators remain invariant under the replacement 
\be
o_\a( \r) \rightarrow \ell^{x_\a}o_\a(\ell \r) \la{sco}
\ee
One can define corresponding scaling  for the other quantities in the theory.

These results enable us to gain very substantial information about critical behavior. The only possibility for $<o_\a( \r)>$ at criticality consistent with \eq{sco} is that this average should be zero for  for $x_\a$ different from zero. 
In an additional, less trivial, example, a correlation between two densities at the critical point has the power-law behavior 
$$
<o_\a(\r) o_\b(\s)>  = C_{\a,\b}/ |r-s|^p
$$
where $C_{\a,\b}$ is an unknown coefficient.  This structure is the only possibility consistent with translational invariance. Now use \eq{sco}. 
This replacement produces no change  if the index, $p$, has the value $p=x_\a +x_\b$.  Thus
\be
<o_\a(\r) o_\b(\s)>  = C_{\a,\b}/ |r-s|^{[x_\a +x_\b]}
\la{critc}
\ee 
Analogous results give all the the scalings proposed by Widom\cite{Widom,WidomII} as well as the correlation function scalings first suggested by  A.Z. Patashinskii and V.L. Pokrovsky\cite{PP}.

\subsubsection{Universality classes\la{Universality} }

To study critical phenomena based upon renormalization transformations, one sets all the relevant combinations of couplings to zero and then does a sufficient number of successive renormalizations so that all the irrelevant combinations have effectively disappeared. All couplings will now remain unchanged under the renormalization. We  thus end up with a  {\em fixed point} independent of the value of all of the irrelevant couplings.   The act of renormalization is a sort of focusing in which many different irrelevant couplings fade away and we end up at a single fixed point representing a whole multi-dimensional continuum of different possible Hamiltonians. These Hamiltonians form what is called a {\em universality class.}
Each Hamiltonian in its class has exactly the same critical point behavior, with not only the  same critical indices, but also the same long-range correlation functions, and the same singular part of the free-energy function.  

The identity among different problems is not just a theoretical artifact. The Ising model, single axis ferromagnets, and the liquid-gas phase transition all show identical critical properties\cite{Lee-Yang,Lee-YangII}.   The renormalization theory makes the fixed points, and thus the universality classes, have properties that vary with dimension. Experiments bear out the predicted universality in the two observable cases: $d=2$ and $d=3\cite{67Review}.$ 

 There are, of course, many different universality classes corresponding to different dimensionalities, to different symmetries of the order parameter, and to different stability properties of the fixed points.   For example, the Ising model, XY model,  and Heisenberg model, respectively having one, two, and three components in their spin vectors, each have different critical fixed points in three dimensions.

\subsubsection{Marginal behavior and critical lines\la{marginal}}
Before leaving this subject, focus once more on the possibility of a marginal behavior.   A marginal operator implies  a coupling with a magnitude that does not vary under renormalization. Continuous variation of this parameter can produce a continuously varying line of critical points, which would otherwise be impossible\cite{Kadanoff-Wegner}.  This kind of critical line is found in the XY model in two dimensions (see \se{XY} below).

\subsection{Qualitative properties\la{qualitative}}

\subsubsection{A new kind of calculation}
The renormalization group method gives a mode of calculation conceptually very different from the older ones. In the old theories, one calculated the properties of the system using the ensemble.  In the new kind of theory, one calculates a renormalized ensemble based upon the previous ensemble.  The old theory called for one step of calculation; the new one calls for a potentially unbounded number of new ensembles.  This new mode of calculation is important, original, and useful. In that sense, the renormalization method is itself a deep and fundamental contribution to knowledge. 

\subsubsection{A new relation between large and small}
The renormalization approach provides a natural explanation of the  extended singularity theorem,  which states that mathematically sharp phase transitions only exist within infinite systems.  

 Furthermore, the fixed-point concept describes a connection between the microscopic properties of the material, i.e. the interactions among its constituent particles and fields, and the behavior of the material on a conceptually infinite length scale.  This connection is surprising and quite beautiful.  The connection also suggests a program of further investigation in which the dependence of the overall properties of the material is studied as a function of changes produced at its far boundaries, or of topological changes in the material itself.   People are just beginning to do this new kind of calculation.
      
\section{Impact of Renormalization Theory}

In condensed-matter applications,renormalization usually bridges scales in the range between the microscopic scales of particle interactions and the macroscopic scales of the laboratory and workplace.     Condensed-matter physics has taken as its task the development of the relation between non-trivial macroscopic theories, of hydrodynamics, elasticity, electrodynamics of materials, strength of materials, etc.  and the non-trivial microscopic theories.   Ideas move smoothly up-and-back between the two kinds of domains. 

In particle physics, the early forms\cite{SP,GML} of the renormalization method were constructed  as an attempt to handle the infinities inherent in all relativistic field theories.  Wilson's work showed that the method could more aptly serve as a bridge between the very different scales that define the different kinds of interactions among the particles arising from various kinds of field theories.  

In addition, many other fields also require a bridge between different scales.  In hydrodynamic turbulence large-scale forces produce small-scale swirls. Dust aggregates into stars and galaxies. Individual economic actors form world-scale markets. Tiny organisms produce huge ecosystems. Repeated action of wind and water grind down mountains. Tremendous earthquakes are produced by small-scale slippage. All of these phenomena require insights that bridge huge length scales, and which  might perhaps be provided by a renormalization approach.   So it is quite natural that the Wilson et. al. phase transition work had as its immediate consequence  an explosion of theoretical activity in diverse fields.

\subsection{Expansion in $\e=4-d$}
In a very early application of the renormalization method,  Wilson and Fisher\cite{Wilson-Fisher} showed how to use the renormalization method by considering the result of varying the number of dimensions, $d$. This began as an expansion of the fixed point for Ising-like systems about four dimensions, using the  momentum-space renormalization scheme described in paper II,  section 7.1\cite{LPKwebsiteII} or from \eq{lagrangian} taken into a continuum spatial formulation.   Four dimensions is special because  the fluctuation-dominated fixed point and the mean field theory fixed point come together at four dimensions.   Near four dimensions, at the fixed point, the Lagrangian density is dominated by quadratic terms in the field variables while the quartic and higher terms determine the fixed-point behavior but are very small.  Thus, with a sufficiently delicate methodology, one can do an expansion about a simple MFT fixed point, quadratic in the fluctuating field. This result, called the {\em $\e$-expansion}, gives an accurate estimate of properties of  both fluctuation-dominated and MFT  fixed points near four dimensions. Furthermore it is an asymptotic expansion in which the lower-order  terms are easily calculable and give accurate estimates of the critical properties of a wide variety of three-dimensional situations.   
 
One important idea permitting this calculation was to consider the spatial dimension of the system to be a continuously variable parameter.   This idea was developed by Michael Fisher in phase transition theory\cite[chapter 4]{Collins}, and then later apparently independently developed\cite{HVa,HVb} to provide part of a proof of renormalizability of particle physics' gauge field theory.

This Wilson-Fisher calculation proved to be very important.  First, in giving estimates of three-dimensional behavior that agreed in a pleasing fashion with series expansions and experiment, it provided a strong indication that the renormalization method was correct.  Second, the $\epsilon$-expansion method was relatively easy to use and to apply to a wide variety of problems.   It required only a little overhead in terms of either learning time or investment in computational facilities and technique.  A wide variety of individuals could join in and share the pleasure of doing realistic calculations of critical properties.    Particle-physics field theorists and condensed-matter theorists could and did participate equally.

\subsection{Gauge field theory}
Soon after the 't Hooft-Veltman proof of renormalizability of gauge theory\cite{HVa,HVb}, the actual renormalization of gauge field theory was carried out by David Gross and Frank Wilczek\cite{GW} and independently by David Politzer\cite{Politzer}. They calculated renormalized couplings in the standard model of particle physics and showed how particle scatterings weaken at higher energies. This property,  called {\em asymptotic freedom},  permits accurate calculations of most of the effects of these strong interactions at higher energies.  
 
The asymptotic freedom calculations were both elegant and important. The results were pretty, compact, and easy to understand. Their importance lies in the fact that the weakening of the interactions at higher energies means that perturbation expansions would converge rapidly at high energies. Thus asymptotic freedom brought within reach accurate calculations of results from the ``standard model'' of particle physics, which could later on be applied to the energy range appropriate for the Large Hadron Collider.  

The relative weakening of strong interactions in going from larger to smaller length scales would continue until the unification scale, when ``weak'' and ``strong'' interactions would attain the same strength.  Further scale reductions would cause strengthening of all these interactions.  This strengthening would indicate that the perturbation theory, and indeed the entire standard model could not serve as an ``ultimate'' theory that might consistently explain all particle physics phenomena.

\subsection{Kondo problem\la{Kondo}}
The Kondo\cite{Kondo}  effect is an anomaly in the electrical resistance of metallic materials with a small admixture of magnetic impurities.  Unlike most other materials, these metals show an electrical resistance that goes through a minimum value at low temperatures.   Jun Kondo's name was attached to this effect after he pointed out the source of this anomaly in the correlated scattering of electrons within the material from the spins on  atomic impurities.  The effects of multiple scatterings of many electrons from a single spin are stored in an interesting fashion in the entanglement of the wave function of that spin with the spin states of the electrons.

Kondo's work explained the basic effect but did not predict  the low-temperature properties of the material because his perturbation expansion  would not converge at low temperatures.      In a notable {\em tour de force} Kenneth Wilson used the connection between quantum mechanics and statistical mechanics to understand this low temperature behavior.     Wilson then solved this quantum problem by numerical application of the renormalization group\cite{Wilson-Kondo}.  Wilson included a very large number of interaction terms in his Hamiltonian and successfully described the  wide range of electronic energies encompassed by this physical effect.  In subsequent calculations, Pavel Wiegmann\cite{Wiegmann} and Natan Andrei\cite{Andrei} provided exact solutions for the Kondo problem that substantially agreed with Wilson's numerical work.

\subsection{Onset of chaos}
One of the interesting discoveries of the twentieth century was {\em chaos}, the tendency of dynamical systems to show complex and never-repeating behavior. The concept of chaos also implies that  very small modifications in initial data can produce arbitrarily large changes in subsequent behavior. Chaos theory gave science a different view of the world from that prevalent in the Nineteenth Century. In this less-deterministic view, despite the overall causality built into classical mechanics and even quantum theory, many important events become, in practice, unpredictable.    

Chaos ideas were developed with the aid of simplified models exemplifying chaotic behavior\cite{Lorentz}.   Probably the simplest example of a dynamic system that can show chaos is the {\em logistic map}, in which a succession of $x$-values, $x_1,x_2,x_3, \cdots$ is produced by the algebraic expression:
\be
x_{j+1} = r x_j (1-x_j).
\la{logistic}
\ee
This  model is generally studied for $r$-values between $0$ and $4$ and $x$-values between $0$ and $1$.  The qualitative  behavior of this dynamical system depends upon the parameter $r$.  For the smallest values of $r$, the model shows an easily predictable behavior. Specifically for $0<r<r_0=2$, any starting value of $x$ will produce a sequence that approaches a specific $r$-dependent value, in particular $x^* = 1-1/r$.  This behavior is  quite orderly.  On the other hand, for larger values of $r$ the model can generate an apparently disorderly sequence of $x$'s.  Ultimately, for  $r=4$, the result is entirely chaotic. At this value of $r$, a typical starting value, $x_0$, will then engender a sequence that cycles arbitrarily close to all the possible values between zero and one.   A starting value just a tiny bit different from this initial choice of $x_0$ will after a while produce a sequence of $x_j$ values completely different from the sequence produced by $x_0$. 

One can ask for what value of $r$ this chaos will first appear, and how it will manifest itself. The answer was given in mathematical detail by Mitchell Feigenbaum\cite{Feigenbaum} using renormalization techniques.  What happens is that for values of $r$ just beyond $r_0$-- in the range $r_0<r<r_1$-- the $x_j$ values will settle down to cyclic behavior, with a cycle of length two: large, small,large, small, ... . Then for slightly larger values of $r$, a cycle of length four stably appears. In a range of $r$'s slightly larger yet there is a cycle of length eight.  With increasing $r_n$'s, cycles of increasing length, $2^{n+1}$, successively appear until at some special, magical, value of $r=r_\infty$  an infinitely long cycle appears.  For larger $r$ than that, one sees a behavior that is complicated and begins to show chaotic properties.  

Perhaps one should ask what all this has to do with the previously discussed work in phase transitions and field theory. The answer is that the phenomena have considerable intellectual overlap in that this period-doubling route to chaos shows both scaling and universality.  The $r_n$- values at which the successive  period-doublings occur show a pattern described by a power-law behavior, 
$$
r_{n+1}-r_n  = \text{constant~} \a^{-n}
$$
for large values of $n$,  as do other properties of the long cycles.    Universality is even more evident.  Not only do different algebraic forms not closely related to the one in \eq{logistic} show the same pattern of period doubling, with for example the same value of $\a$, but so do experimental observations like those given by non-linear electrical circuits\cite{TPJ} or hydrodynamical approaches to turbulence\cite{multifractal}.          

However none of these parallels to Wilson's work was evident when Feigen{-}baum\cite{Feigenbaum} started his empirical study of \eq{logistic}.  Much of the previous work on chaos had emphasized the complexity and richness of the patterns produced by equations like this. Indeed well-developed chaos is quite, well,  {\em chaotic}. However, the onset of chaos can be simple and can show a mathematical structure that can be well-described by renormalization techniques.   Feigenbaum's work was the first analysis of onset along these lines and was followed up by a rich collection of other mathematical and physical studies that described this route to chaos in more detail, or developed the properties of other routes to chaos. All of this was based upon renormalization arguments.  
\subsection{Real space renormalization}
As already noted, much of the most accurate and complete work on three-dimensional critical phenomena could be done with the aid of $\e$-expansion methods.   However, these expansions do not converge rapidly enough to be maximally useful for two-dimensional problems.  Instead, the method of choice in two dimensions is to make approximations directly upon block transformations like the one in \fig{block}, as was first done by Th. Niemeyer and J. M. J. van Leeuwen\cite{NL}.  Approximations derived in this way have been used very extensively\cite{LPK00},\cite[Niemeyer and van Leeuwen, volume 6, 1976]{DGL} and often appear to be quite accurate, but it is hard to estimate the size of their errors.

These renormalization techniques seem to work quite well despite the fact that all approximate renormalization analysis has, so far, not yielded to a justification by a rigorous mathematical analysis. 
    
\subsection{XY model\la{XY}}    
 In 1973, Michael Kosterlitz and David Thouless\cite{KT} produced an approximate solution that described the phases and phase transitions in a model involving fixed-length magnetic spins, $\mathbf{s}=(s_1,s_2)=(\cos \theta, \sin \theta)$,   capable of pointing in any direction in a two-dimensional space.  Their calculation  followed in part the earlier work of Vadim L'vovich Berezinskii\cite{BerezinskiiI,BerezinskiiII}.  Work on this model is important in itself, but it is also illustrative of the many substantial advances in critical phenomena that appeared immediately after the development of  renormalization theory.  
 
 The XY studies were notable in several different regards:
\begin{itemize}
\item  The solution to this model closed a problem left unexplained by mean field theory.  According to MFT, the XY model should have both a phase transition and long-range order.  But as shown by a theorem of David Mermin and Herbert Wagner\cite{MW} and also by Pierre Hohenberg\cite{Hohenberg},  no long-range order is possible for vector magnets in two dimensions.  When a range of directions is equally possible, there cannot be a two-dimensional ordered state in which the magnetization vector points in a specified direction. In this work of the early 1970s,  Berezinskii, Kosterlitz, and Thouless  show that there is a line of phase transitions for two-dimensional magnetization vectors in two-dimensional space..  Instead of an ordered state, the model has a magnetization correlation function showing an algebraic behavior, with a critical index that depends on the strength of the coupling. (Specifically the critical index called $x$ in \eq{corr} varies continuously with the strength of the coupling.) This behavior was entirely unexpected, but it is supported by a solid renormalization argument\cite{Joseetal}.  
\item
This XY model is an important illustration of universality.    Not only does this model describe a particular kind of magnet, it also equally well describes a superfluid-to-normal phase transition.  The superfluid is described by a complex wave function and the two-component vector of the magnet could serve as a surrogate for the real and imaginary parts of a complex wave function.  In addition this same universality class includes the behavior of a fluid-to-solid transition in two dimensions.  The two-dimensional solid is described by a local orientation and that orientation can become ordered (the solid phase) or disordered (the fluid phase).  Yet one more member of this rich universality class is the transition between a plasma containing charged particles and long-range interactions and an ordinary fluid in which the charges are screened and the effective interactions are short-ranged.
\item 
 The excitations in the critically ordered state of this model are unique and interesting.  One kind of excitation, called a vortex, involves in a essential way the topology of a two dimensional vector.   This excitation depicted in \fig{vortex} is a swirl extending out to infinity.  The crucial characteristic of the two-dimensional vector defining the local direction of the swirl is that when it is rotated through $360^\circ$ it returns to its starting point.  This characteristic makes the mismatch in direction among neighboring spins very small except at the singularity at the center of the vortex.   It is this property that makes the vortex excitation a characteristic part of the state generated by the XY model.

Topological excitations can be observed in many other statistical systems, but the XY model provided the first example and the one that is easiest to understand.
\end{itemize}

     \begin{figure}
%\begin{multicols}{2}
\includegraphics[height=4 cm ]{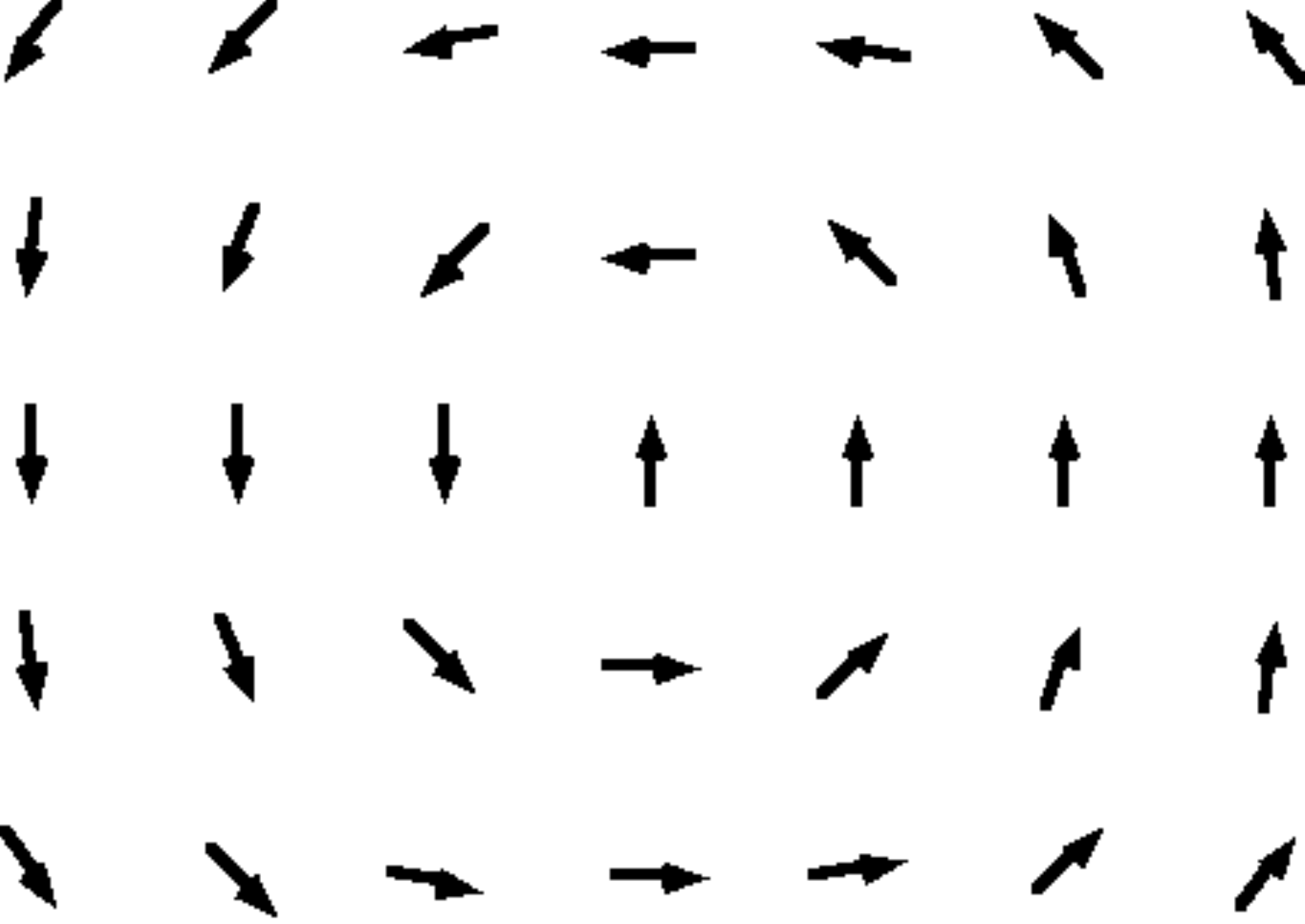}
\hskip 45 pt
\includegraphics[height=4.5cm ]{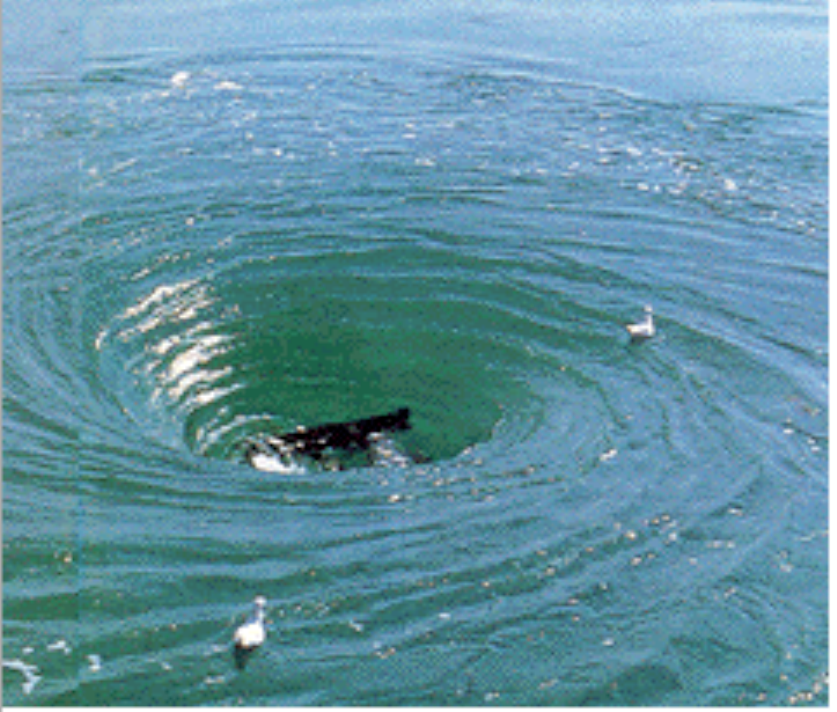}
\caption{Vortices.  The left-hand  panel is a cartoon view of a vortex  excitation in the XY model.  The right hand panel is a hydrodynamic swirl with floating birds to set the scale.   }  
\la{vortex}
%\end{multicols}%{2}
\end{figure}

For all of these reasons, the XY model is quite special and specially interesting. This model, along with the other applications mentioned in this chapter all were developed in the years immediately after Wilson's extension of renormalization concepts and can be considered direct outcomes of that extension.

\subsection{Conformal field theory\la{CFT}}
Even before Wilson's work, Polyakov\cite{Polyakov} developed a method called conformal field theory (CFT) that added substantially to our knowledge of phase transitions and field theories.   (For later work of Polyakov and coworkers on this problem see \cite{BPZ}.)  This kind of field theory applies at criticality, or in particle physics language, when all the particles defined by the field theory have mass zero.  Conformal field theory (also called CFT) goes one step beyond the obvious geometrical symmetries of the previous critical-point theories.  We know that the theory is going to have a symmetries based upon translational invariance ($\r \rightarrow \r +\mathbf{a}$), rotational invariance, and scale symmetry ($\r \rightarrow \ell~ \r $).  Polyakov noted that most field theories with these symmetries also have an additional symmetry involving the point at spatial infinity.  In cases in which this is just an ordinary point just like any other, the transformation that takes zero into infinity  has the property that,  if one changes the most basic local operator via 
\be
o_\a( \r) \rightarrow    r^{-2x_\a}  o_\a(\r/r^2),   \la{conformal}
\ee   
this change will leave all correlation functions invariant. This condition then provides important constraints upon correlation functions. For example, \eq{corr} for the correlation between two local operators is not in general consistent with the conformal symmetry. The only way a non-zero correlation of two basic operators can be invariant the transformation of \eq{conformal} is for these operators to have the same $x_\a$-value. Otherwise the correlation function will vanish.  

In a major triumph,  Daniel Friedan, Zongan Qiu, and Stephen Shenker\cite{FQS}  used conformal symmetry to obtain the critical indices of  the most important two-dimensional models of phase transitions.  So far people have not succeeded in extending this work to higher dimensions.

\subsection{Critical dynamics\la{Dynamics}}
The major importance of fluctuations in phase transitions is part of a broader thread of intellectual activity in the post World War II period. The content of the thread is that a wide variety of dynamical processes tend to produce power-law behaviors, which may then be explained by large-scale fluctuation processes.    This thread arose independently of renormalization group activities\cite{Frisch,K62,Benoit,Feder} but gained considerable support from the intellectual strength of phase transition work and the renormalization group. 

I cannot fully describe this kind of work here.  The field is too varied and rich. However, I can list the kinds of problems in which system-wide fluctuations are likely to dominate some portion of the behavior.  This list will give some rough indication of the reach of fluctuation ideas beyond phase transitions.  

One finds among the major applications of system wide fluctuations:
\begin{itemize}
\item In hydrodynamics, turbulence\cite{Frisch,K62}
\item In biology, the formation of different cell types\cite{Kauffman}
\item In paleontology, mass extinctions
\item In geosciences, earthquakes
\item In geosciences, shapes of natural objects including as described in studies of fractal behavior\cite{Benoit}
\item In material sciences, the transition from conductors to insulators.
\end{itemize}

\subsection{Financial models}  Economic theory, as first worked out by Adam Smith and then further expounded by Paul Samuelson, has an equilibrium model that treats money in much the same manner as Maxwell and Boltzmann treated energy in their development of statistical mechanics.    Generations of economists have refined, developed, and applied this model.  One of the highest-impact results of this kind of thinking came in the  
1973 work of Fischer Black and Myron Scholes\cite{BlackScholes}, further expounded and explained by Robert C. Merton\cite{Merton}.   They put forward a theory of the pricing of derivative securities, i.e. securities that gave one the right to buy or sell other financial instruments at a future time under specified conditions. From this theory, one could deduce a model predicting a ``rational'' price for a derivative security.   This Black-Scholes model had some of the same content as the mean field theory models used to describe phase transitions.  Specifically, the financial model calculated the future value of a derivative security by taking averages in an assumed statistical market environment.  As with MFT, fluctuations were treated approximately and correlations among fluctuations that might exist over the entire market were ignored.

More broadly, economic theories,  as they are used today, have much of the content of MFTs.  Averages are included, but major fluctuations and large-scale correlations are usually neglected.

Prices in the market for derivative securities have in the main followed the theories of Black and Scholes.  In fact, an entire industry has been constructed using their conceptualization to form a rational basis for pricing and trading these securities.   However,   the theory is not perfect.  The theory correctly predicts probabilities for prices close to the average.  This agreement most likely is in part due to the virtues of the theory, and in part because the majority of the speculators who formed the market worked with and believed in the Black-Scholes theory. Financial markets show well-known departures from this prediction.   Specifically the pattern of probabilities exhibit what are termed ``fat tails'' in that extreme prices are seen empirically to occur much more frequently than predicted by the theory\cite{econophysics}.    

One of the leading firms using the theory to profit from the market was a hedge fund with the name {\em Long Term Capital Management}.  The board of Long Term Capital Management included both Scholes and Merton. For several years, the firm showed a 40\% per year profit on investments. However,   in 1997 and 1998 the firm quite spectacularly went bankrupt and was bailed out by a large portion of the financial industry.  The bankruptcy was caused in part by abnormally large excursions in financial markets, excursions that apparently were not included in the modeling.   

In 2008 the entire market for derivative securities  showed a similar catastrophe, one that was mitigated for the industry by a public bailout.  It is not known  to what extent the exclusion of fluctuations from financial models and the use of MFT ideas contributed to the result.  However, many observers believe that modeling of entire markets and entire economies would benefit from a more explicit inclusion of large-scale fluctuations, and perhaps also of the effect of human emotions.

\section*{Acknowledgments} Some of the material in this review   was first prepared for a talk I gave at the Royal Netherlands Academy of Arts and Sciences in 2006.  Material in this paper appeared in a talk at the 2009 Seven Pines meeting on the Philosophy of Physics\cite{I} under the title ``More is the Same,   Less is the Same, too;  Mean Field Theories and Renormalization.''  These talks have appeared on my web site\cite{LPKwebsiteII} since then.   This Seven Pines meeting was generously sponsored by Lee Gohlike. 

This work was was started with support from the University of Chicago MRSEC program under NSF grant number DMR0213745.  It was completed with support from the present MRSEC grant, number 0820054, and  during  visits to the Perimeter Institute,  which is supported by the Government of Canada through Industry Canada and by the Province of Ontario through the Ministry of Research and Innovation.

I had useful discussions related to this paper with  Tom Witten,  E. G. D. Cohen, Gloria Lubkin,  Michael Fisher,  Emanuel Derman, Gene Mazenko, Hans van Leeuwen, Wendy Zhang, Michael Goodkin, Sidney Nagel, and Jon Rosner.

\bibliographystyle{plain}
\bibliography{philo,Part,LPK}
\end{document}